\documentclass[pra,aps,twocolumn,showpacs,10pt]{revtex4-1}
\usepackage{amsmath,amssymb,graphicx,bm,color,ulem}
\usepackage{amsfonts,amsthm}

\begin{document}

\title{Exploding paraxial beams, vortex beams, and cylindrical beams of light with finite power in linear media, and their enhanced longitudinal field}

\author{Miguel A. Porras}

\affiliation{Grupo de Sistemas Complejos, ETSIME, Universidad Polit\'ecnica de Madrid, Rios Rosas 21, 28003 Madrid, Spain}

\begin{abstract}
We present a set of paraxial light beams with cylindrical symmetry, smooth and localized transversal profile carrying finite power, that develop intensity singularities when they are focused in a linear medium, such as vacuum. They include beams with orbital angular momentum and with radial polarization, in which case they develop punctual phase and polarization singularities surrounded by infinitely bright rings, along with singular longitudinal fields. In practice, these effects are manifested in focal intensities and spot sizes, vortex bright ring intensities and radii, and strengths of the longitudinal field, that strongly change with the lens aperture radius. Continuous control of these focal properties is thus exercised without changing the light incident on the lens, with substantially the same collected power, and while maintaining paraxial focusing conditions. As solutions of the Schr\"odinger equation, these exploding beams have analogues in other areas of physics where this equation is the fundamental dynamical model.
\end{abstract}

%\pacs{32.80.Wr; 42.65.Tg; 05.45.Yv}

\maketitle

\section{Introduction}

Inspired by what happens to some wave functions in quantum mechanics \cite{PERES}, Aiello has recently introduced a class of paraxial light beam with a localized transversal profile and finite power that develops a singularity when its is ideally focused \cite{AIELLO}. Real, apertured versions of these beams have subsequently been demonstrated in experiments \cite{AIELLO2}. These beams, even if apertured, outperform standard Gaussian beams of similar intensity and power in terms of focal intensity and resolution. In a sense, these beams are able to reproduce with a finite amount of power the focusing behavior of plane waves, though their finite-aperture realizations do not exceed the diffraction limit \cite{AIELLO}. The electric field of these beams is factorized in the coordinates $x$ and $y$ in the transversal plane, i. e., it has a rectangular geometry, which greatly facilitates its analytical treatment.

In many experimental settings, however, cylindrical symmetry is advisable, if not mandatory; in particular, lenses, whose aperture plays a crucial role in the behavior of these beams, are most often circular. In this paper we describe cylindrical symmetric beams with finite power and similar exploding behavior in their intensity when they are focused. Considering beams with cylindrical geometry, we can also examine other types of beams that are of great interest today. We describe exploding beams with orbital angular momentum, and exploding radially polarized beams, both with finite power. They form a punctual dark vortex surrounded by an infinitely bright ring, accompanied by an infinitely strong longitudinal component.

Their apertured versions approach the above ideal behavior as the lens aperture is increased, which offers a practical way to control these properties without changing the illuminating beam, including the collected power, e. g., to enhance the amplitude of the axial component up to $66$ percent of that of the transverse component under paraxial conditions. These exploding beams, vortex beams and radially polarized beams with large, but paraxial apertures, greatly outclass the performance of standard illuminating fields such as Gaussian or Laguerre-Gaussian beams of similar power and intensity. As their Cartesian counterparts, exploding beams of finite aperture do not beat the diffraction limit for plane wave illumination, but approach it closely. The practical advantages of exploding versus uniform illumination for controlling the properties of the focused light is discussed at the end of this paper.

Of course, the exploding behavior of these light beams has nothing to do with blow-up light beams in self-focusing nonlinear media \cite{KIVSHAR,FIBICH}. Yet, they are somewhat similar in that they both appear in paraxial fields, are dissolved when more precise models are considered, and their mathematical existence has physical repercussions. The results presented here can straightforwardly extended to waves in other areas of physics such as probability waves in quantum mechanics, or mechanical waves in acoustics, and to matter waves and electron beams.

\section{Ideal exploding beams and vortex beams}

In the Fresnel regime, the focused electric field $\psi(x,y,z) e^{i(kz-\omega t)}$ of frequency $\omega$ and propagation constant $k$ (e. g., $k=\omega/c$ in vacuum), and the field $\psi(x,y) e^{-i\omega t}$ at the input plane of a thin lens of focal length $f$, are related by
\begin{equation}\label{FRESNEL}
\psi(x,y,z) = \frac{k}{2\pi i z}\int\!\! dx' dy'\psi(x',y',0)e^{\frac{ik}{2z}[(x-x')^2+(y-y')^2]}\,,
\end{equation}
where $\psi(x,y,0)=\psi(x,y)e^{-ik(x^2+y^2)/2f}$, and the integral extends over the aperture of the lens. We are interested in an illuminating field of the form $\psi(x,y)=\psi(r)e^{is\varphi}$, where $(r,\varphi)$ are polar coordinates in the transversal plane, and $s=0, \pm 1, \pm 2, \dots$ i. e., in light beams with revolution symmetry about the propagation axis without and with orbital angular momentum associated with vortex of topological charge $s$. For these fields, Fresnel integral in (\ref{FRESNEL}) can more conveniently be written as
\begin{eqnarray}\label{FRESNELR}
\psi(r,z)e^{is\varphi} &=& \frac{ke^{is\varphi}}{i^{|s|+1}z} e^{\frac{ikr^2}{2z}}\int_0^R dr' r' \psi(r') \nonumber \\
 &\times& e^{\frac{-ikr^{\prime 2}}{2}\left(\frac{1}{f}-\frac{1}{z}\right)} J_{|s|}\left(\frac{k}{z}r r'\right)\,,
\end{eqnarray}
where $J_n(\cdot)$ is the Bessel function of the first kind and order $n$ \cite{GRADS}, and $R$ is the radius of the lens aperture.

\begin{figure}[!]
\begin{center}
\includegraphics*[height=3.75cm]{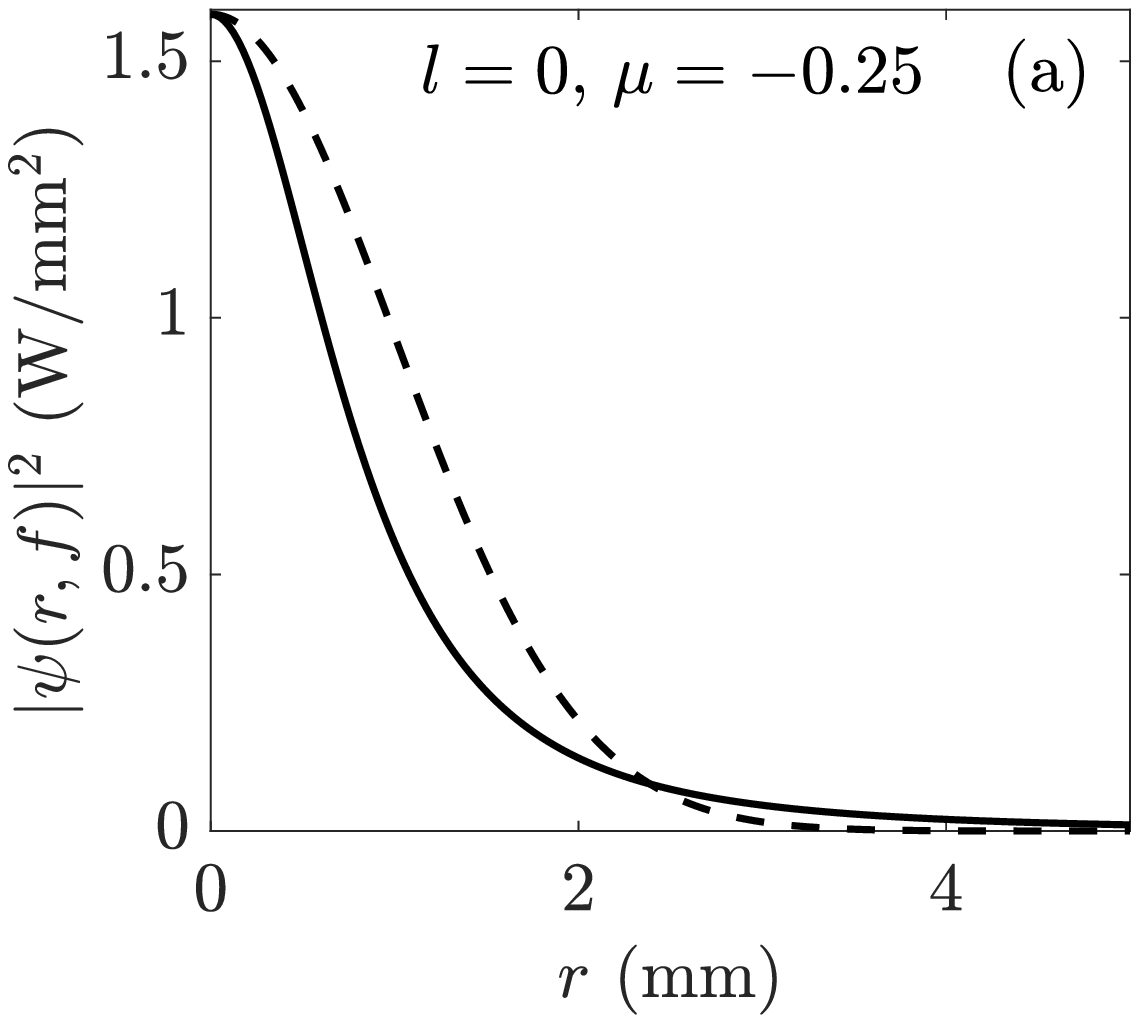}\includegraphics*[height=3.75cm]{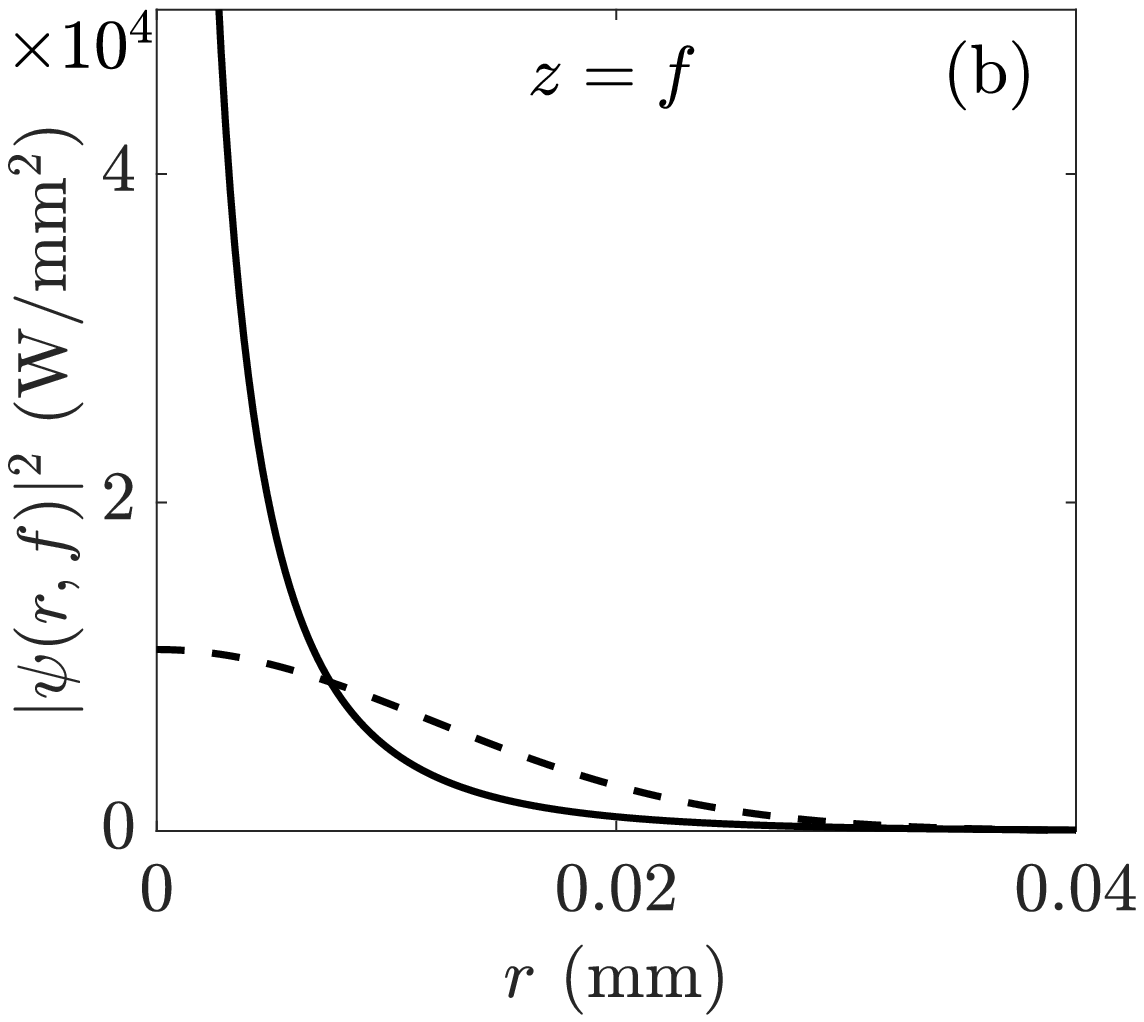}
\includegraphics*[height=3.75cm]{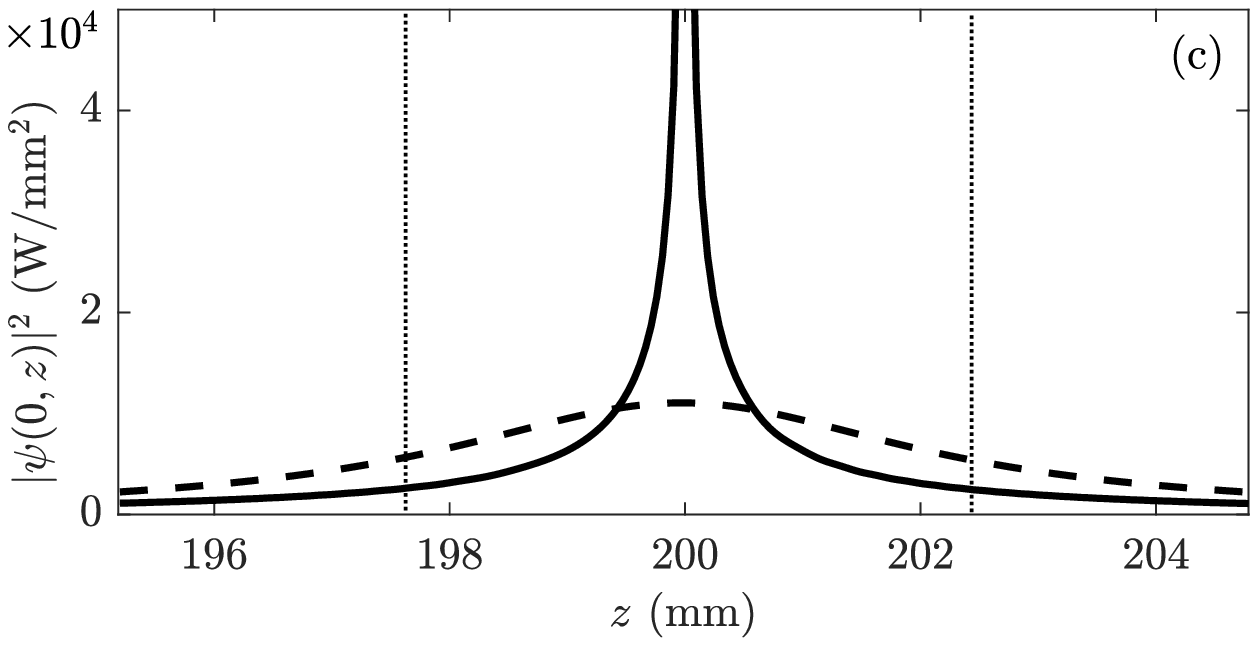}
\includegraphics*[height=3.75cm]{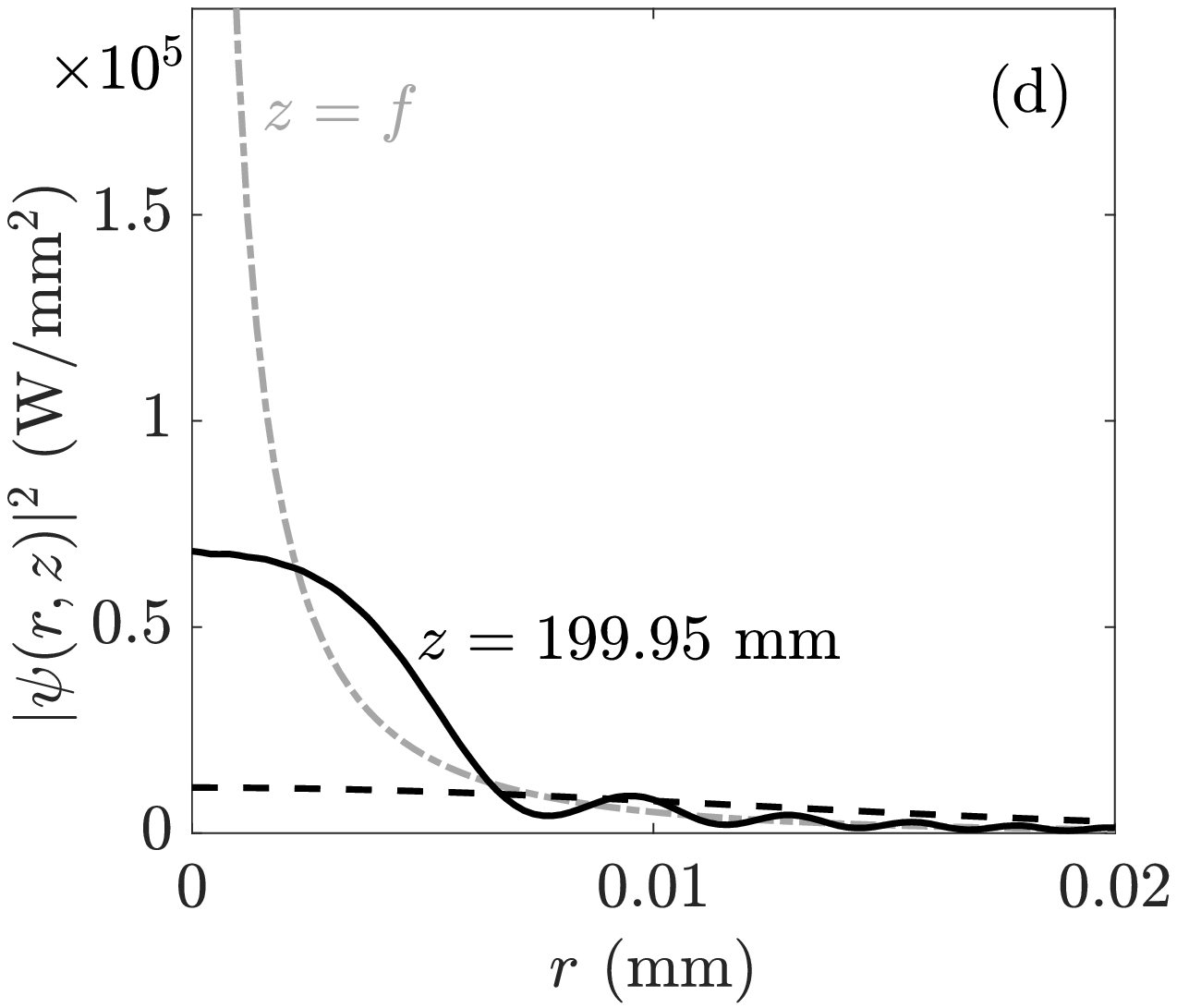}\includegraphics*[height=3.75cm]{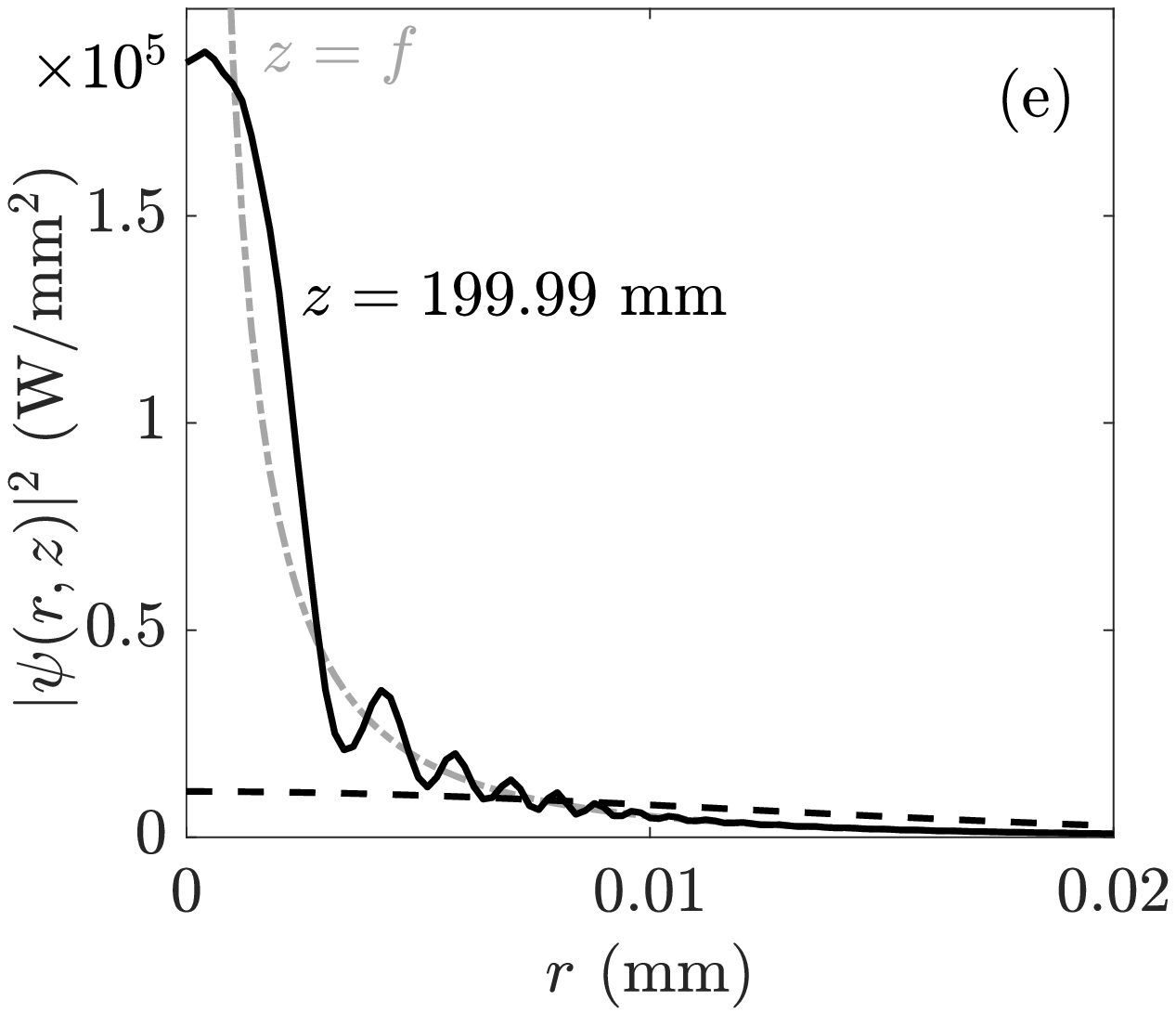}
\end{center}
\caption{\label{Fig1} (a) Intensity $|\psi(r)|^2$ of the exploding illumination in (\ref{ILU}) with power $P=10$ W, $\sigma= 1$ mm, $s=0$, and $\mu=-0.25$. (b) Intensity of the unapertured exploded field at the focal plane as given by (\ref{FOCAL}) when $\omega=2.5$ rad/fs$^{-1}$, $k = \omega/c= 8.33$ mm$^{-1}$, and $f=200$ mm (solid). (c) Peak intensity at $r=0$, numerically evaluated from (\ref{FRESNELR}) with (\ref{ILU}), versus $z$ about the focus. (d) and (e) Radial profiles at planes close to the focal plane, evaluated from (\ref{FRESNELR}) with (\ref{ILU}). For comparison, in all plots the dashed curves represent the same quantities for Gaussian illumination of the same power and peak intensity. The vertical lines in (c) delimit the focal region of the focused Gaussian beam.}
\end{figure}

\begin{figure}[!]
\begin{center}
\includegraphics*[height=3.75cm]{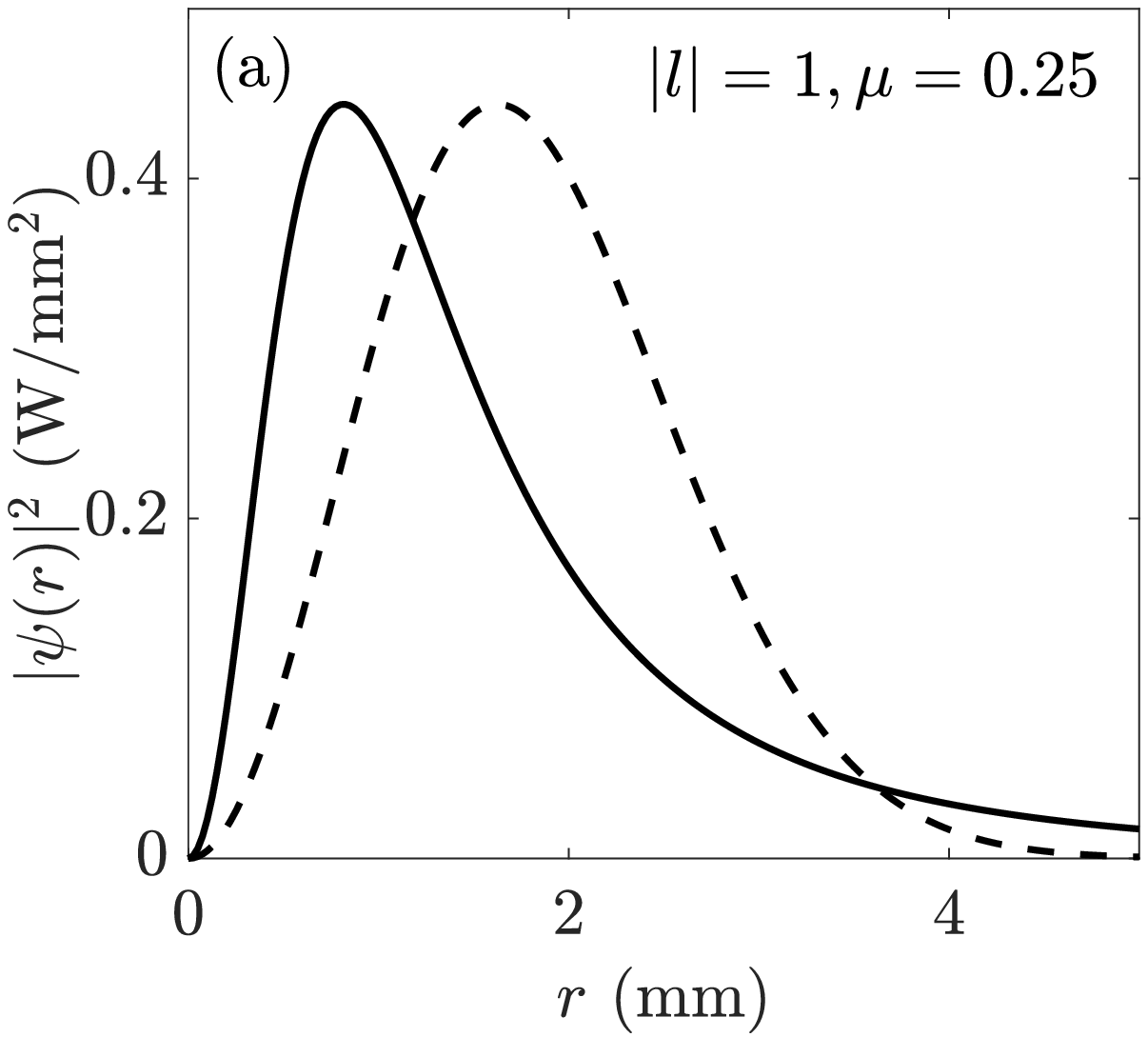}\includegraphics*[height=3.75cm]{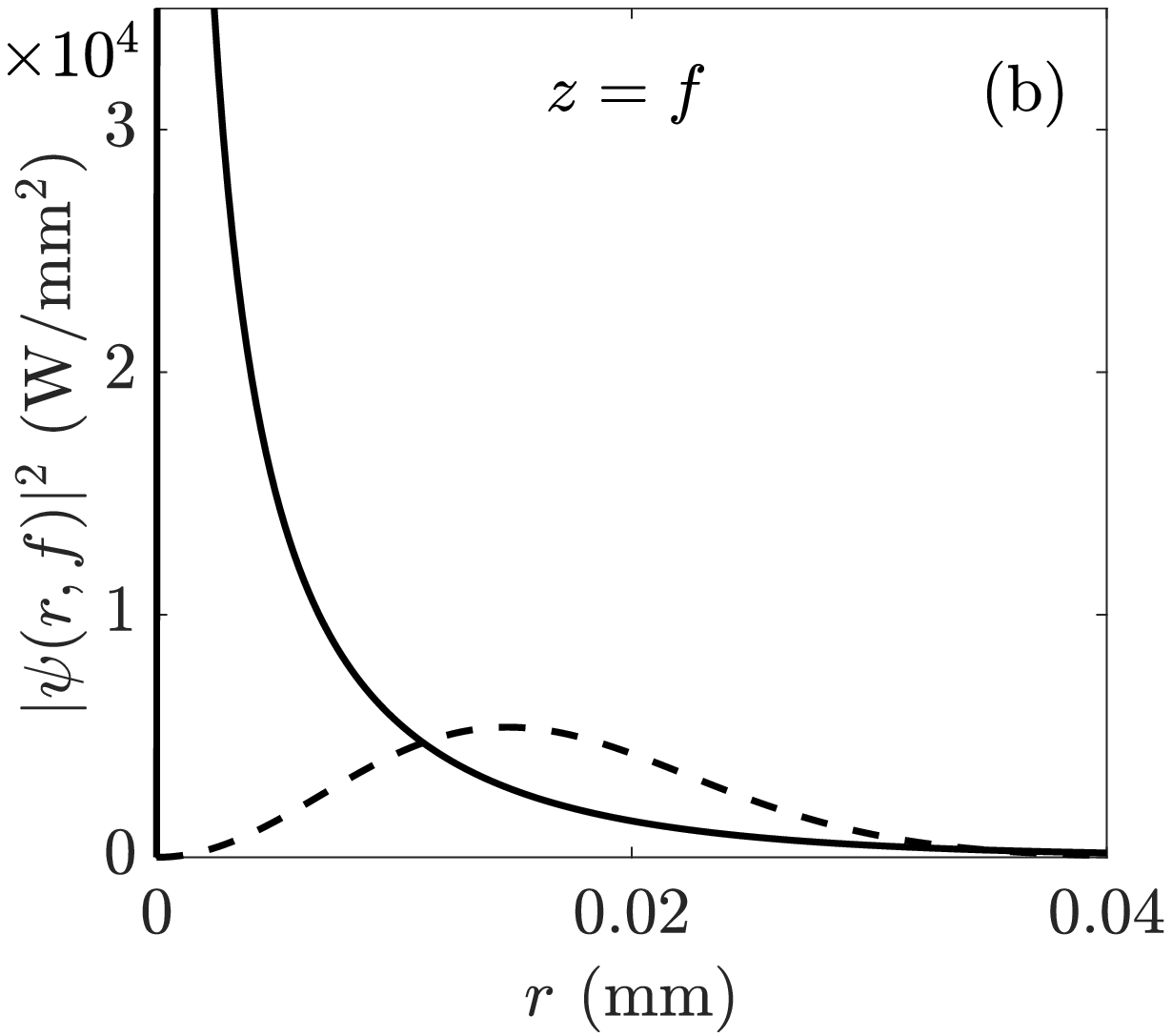}\\
\hspace*{0.35cm}\includegraphics*[width=8.4cm]{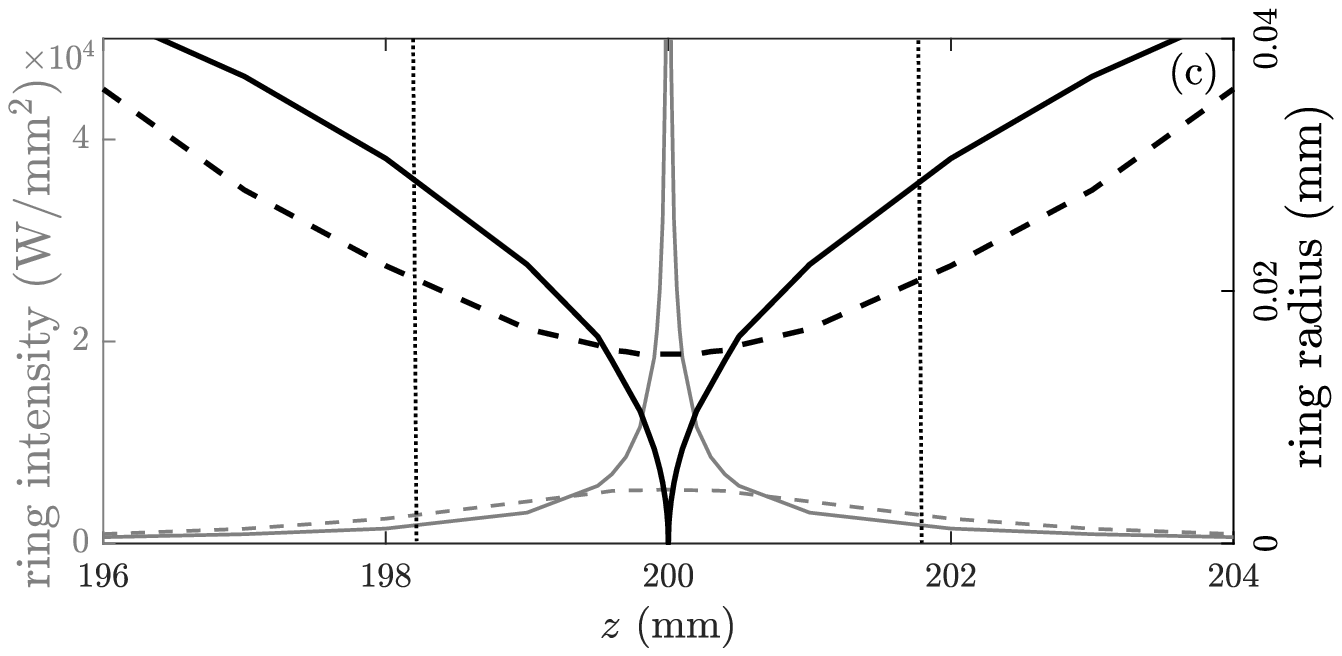}
\includegraphics*[height=3.75cm]{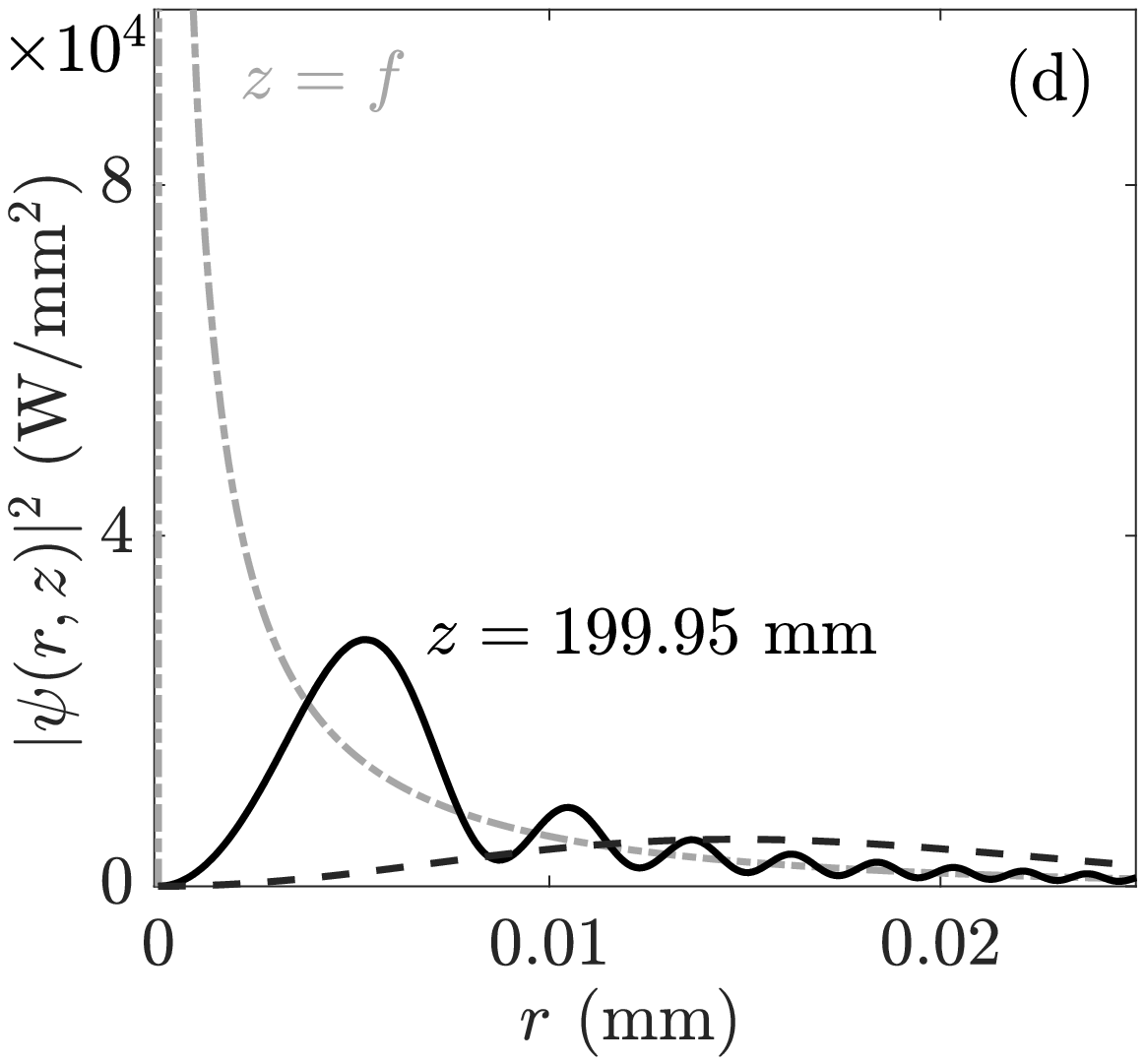}\includegraphics*[height=3.75cm]{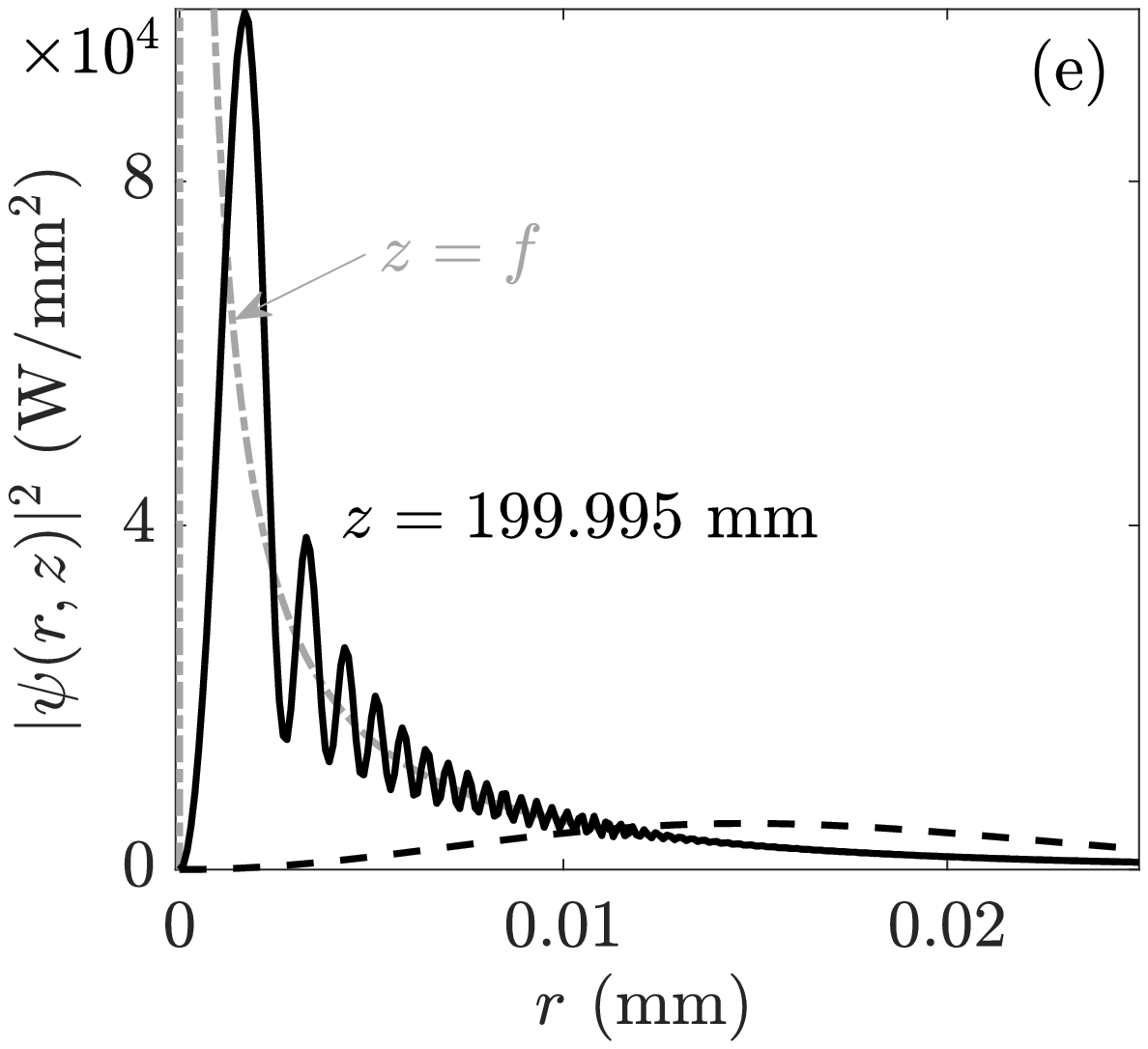}
\end{center}
\caption{\label{Fig2} (a) Intensity $|\psi(r)|^2$ of the exploding illumination in (\ref{ILU}) with power $P=10$ W, $\sigma= 1$ mm, $s=1$, and $\mu=0.25$. (b) Intensity of the unapertured exploded field at the focal plane as given by (\ref{FOCAL}) when $\omega=2.5$ rad/fs$^{-1}$, $k = \omega/c= 8.33$ mm$^{-1}$ (vacuum), and $f=200$ mm. (c) Gray curve: Peak intensity at the maximum of the bright ring surrounding the vortex, numerically evaluated from (\ref{FRESNELR}) with (\ref{ILU}), versus $z$ about the focal point. Black curve: radius of the maximum of the bright ring versus $z$. (d) and (e) Radial profiles at planes close to the focal plane evaluated from (\ref{FRESNELR}) with (\ref{ILU}). For comparison, in all plots the dashed curves represent the same quantities for Laguerre-Gauss illumination with $s=1$ and zero radial order, $\psi(r) \propto r\exp(-r^2/a^2)e^{i\varphi}$, having the same power and peak intensity.}
\end{figure}

We consider the illuminating, collimated field with transversal distribution of amplitude
\begin{equation}\label{ILU}
\psi(r) = \sqrt{\frac{P}{A}}\frac{(r/\sigma)^{|s|}}{(1+r^2/\sigma^2 )^{\mu +1}}\,,
\end{equation}
where the length $\sigma$ scales the field transversally, and $P$ and $A$ are constants to be conveniently fixed. With the real number $\mu$ satisfying $\mu>(|s|-1)/2$, the power of the illuminating field in the entire transversal plane, $2\pi\int_0^\infty dr r |\psi(r)|^2$, is finite, and therefore this field is in principle physically realizable. With the choice $A= \pi\sigma^2 \Gamma(|s|+1)\Gamma(2\mu+1 -|s|)/\Gamma(2\mu+2)$, where $\Gamma(\cdot)$ is the Gamma function, the beam power is just the constant $P$ appearing in (\ref{ILU}). Examples of (\ref{ILU}) with $s=0$ and $|s|=1$ are depicted as solid curves in Figs. \ref{Fig1}(a) and \ref{Fig2}(a).

At the focal plane, $z=f$, the integral in (\ref{FRESNELR}) with (\ref{ILU}) can be carried out analytically (see 6.565.4 in Ref. \cite{GRADS}) when neglecting the finiteness of the lens aperture, i. e., with $R=\infty$:
\begin{eqnarray} \label{FOCAL}
\psi(r,f)e^{is\varphi} &=& \sqrt{\frac{P}{A}}\frac{k}{i^{|s|+1} f}\frac{\sigma^2}{2^\mu\Gamma(\mu+1)} \nonumber \\ &\times& e^{\frac{ikr^2}{2f}} \left(\frac{k}{f}\sigma r\right)^\mu K_{|s|-\mu}\left(\frac{k}{f}\sigma r\right) e^{is\varphi} \,,
\end{eqnarray}
for $r>0$, where $ K_{\nu}(\cdot)$ is the modified Bessel function of the second kind and order $\nu$ \cite{GRADS}. Using that $K_{-\nu}(s)=K_{\nu}(s)$, and the asymptotic behavior $K_\nu(\alpha)\simeq (1/2)\Gamma(\nu) (\alpha/2)^{-\nu}$ for $\nu >0$ for small argument \cite{GRADS}, (\ref{FOCAL}) is seen to approach infinity for $r\rightarrow 0$ when $\mu<|s|/2$. For these values of $\mu$, the vortex-less beam ($s=0$) reaches infinitely large amplitude at the beam center $r=0$, and the vortex beam ($s\neq 0$) vanishes at $r=0$, since $J_{|s|}(0)=0$ for $s\neq 0$, i.e., has an punctual vortex surrounded by an infinitely bright ring. In short, the unapertured focused field of illumination in (\ref{ILU}) carries finite power and produces infinitely intense field approaching the focal point if
\begin{equation}\label{COND}
\frac{|s|-1}{2} <\mu < \frac{|s|}{2}\,,
\end{equation}
and this unbounded field surrounds an infinitely narrow, dark vortex when $s\neq 0$. The singularity originates from the small but persistent tails of (\ref{ILU}) at large radius that are superposed with uniform phases at the focal point, whose contribution to the beam power is nevertheless finite and small.

We point out that the family of light beams with amplitude $\psi(r)e^{is\varphi}e^{-ir^2/2f}$ in (\ref{ILU}) at $z=0$ and the singular field $\psi(r,f)e^{is\varphi}$ in (\ref{FOCAL}) at $z=f$ are solutions of the Sch\"odinger equation
\begin{equation}
\frac{\partial\psi}{\partial z} = \frac{i}{2k}\Delta_\perp \psi\,,
\end{equation}
where $\Delta_\perp= \partial^2/\partial_x^2 + \partial^2/\partial y^2$, and as such these beams can be directly translated to other areas of physics where the Schr\"odinger equation is the dynamical model, e. g., in the dynamics of free electron wave packets in quantum mechanics \cite{BLIOKH}.

Figures \ref{Fig1} (a) and (b) compare the exploding illuminating field in (\ref{ILU}) at $z=0$ and exploded field in (\ref{FOCAL}) at $z=f$ for $s=0$ and $\mu=-0.25$ (solid curves) with illuminating Gaussian beam of the same power and peak intensity and its focused field (dashed  curves). Figures \ref{Fig1}(c), (d) and (e) show peak intensities at $r=0$ versus $z$ and radial profiles at values of $z$ close to $f$ illustrating how the beam profile approaches the singular profile in (\ref{FOCAL}), compared to the same quantities for the Gaussian illumination (dashed curves). The singularity is only formed at the focal plane because the fast oscillations $\exp\left[\frac{-ikr^{\prime 2}}{2}\left(\frac{1}{f}-\frac{1}{z}\right)\right]$ in (\ref{FRESNELR}) out of focus makes the integral to converge to finite values. Compared to the Gaussian illumination, the singularity develops explosively in a tiny axial region of the standard depth of focus [vertical lines in (c)].

Thus, as already pointed out in \cite{AIELLO} and \cite{AIELLO2}, the exploding profile in (\ref{ILU}) reproduces, to a certain extent, what happens when a plane wave of infinite lateral extent and power is ideally focused, namely, both focused fields result in an infinitely bright point at the focus, with the substantial difference that the exploding profile carries finite power, and is therefore physically realizable, in the same sense that a standard Gaussian beam is. This is the cylindrically symmetric counterpart of the concentrating beam factorized in $x$ and $y$ described and realized in \cite{AIELLO} and \cite{AIELLO2}.

Also with cylindrical symmetric intensity, Figs. \ref{Fig2}(a) and (b) show illuminating exploding and exploded transversal profiles of vortex beams with orbital angular mementum given by (\ref{ILU}) and (\ref{FOCAL}) with $s=1$ and $\mu=0.25$ (solid curves), compared to illuminating Laguerre-Gauss beam with $s=1$ (and zero radial order) having the same power and peak intensity, and its focused field (dashed curves). The peak intensity at the radius of the bright ring surrounding the vortex grows up explosively with $z$ to infinity at $f$, as seen in Fig. \ref{Fig2}(c) (solid gray curve), and the same time that the bright ring shrinks acceleratedly down to zero to form a punctual vortex at $f$, as also seen in Fig. \ref{Fig2}(c) (solid black curve). For comparison, Fig. \ref{Fig2}(c) also shows the peak intensity and radius of the bright ring for the Laguerre-Gauss illumination (dashed curves). The ripples in the explosively growing and shrinking radial profiles at planes close to the focal plane, observed in Fig. \ref{Fig2}(d) and (e), disappear in the smoother, singular profile at the focal plane.

\section{Observable effects of the exploding behavior}

\begin{figure*}[!]
\begin{center}
\includegraphics*[height=3.75cm]{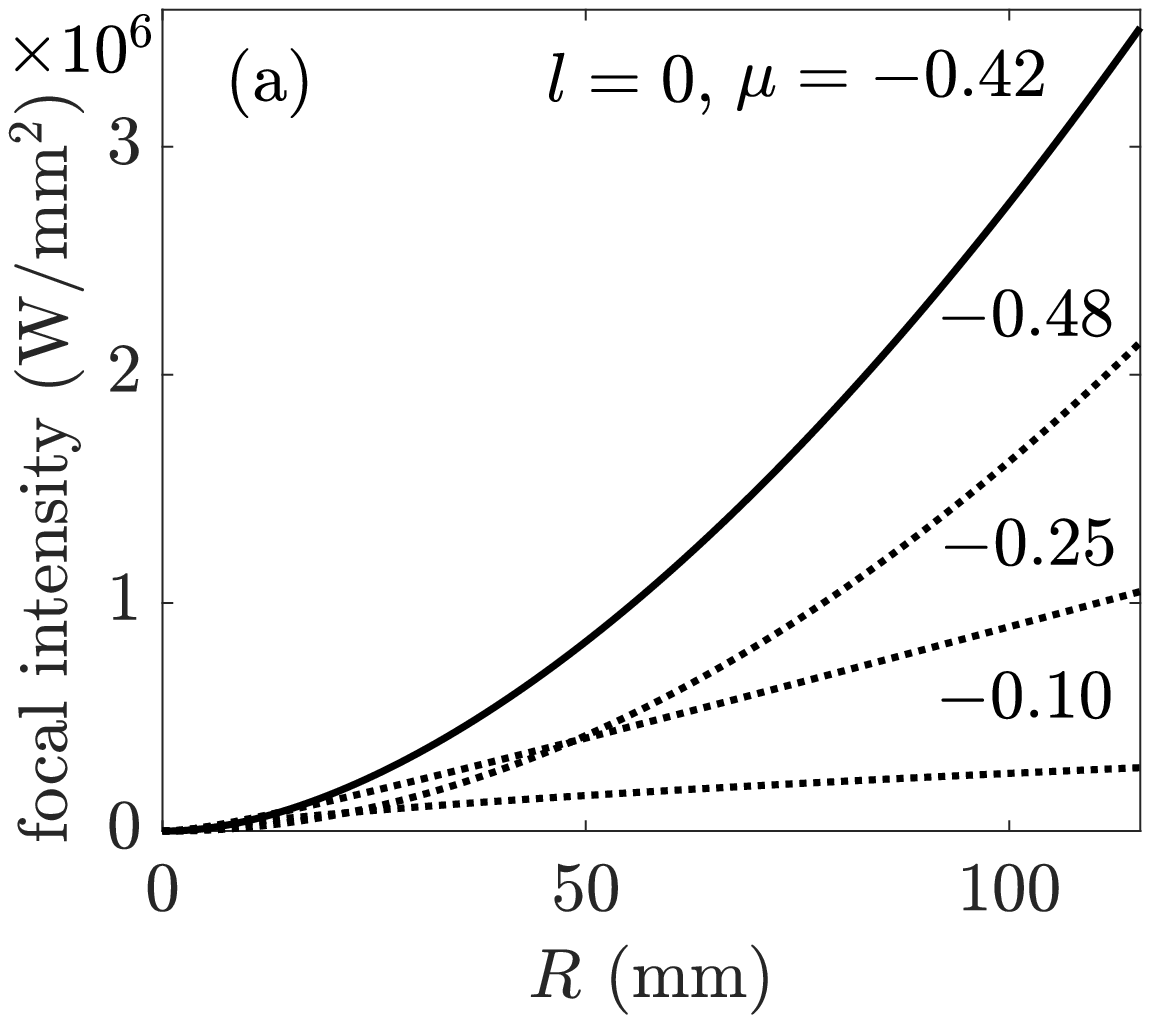}\includegraphics*[height=3.75cm]{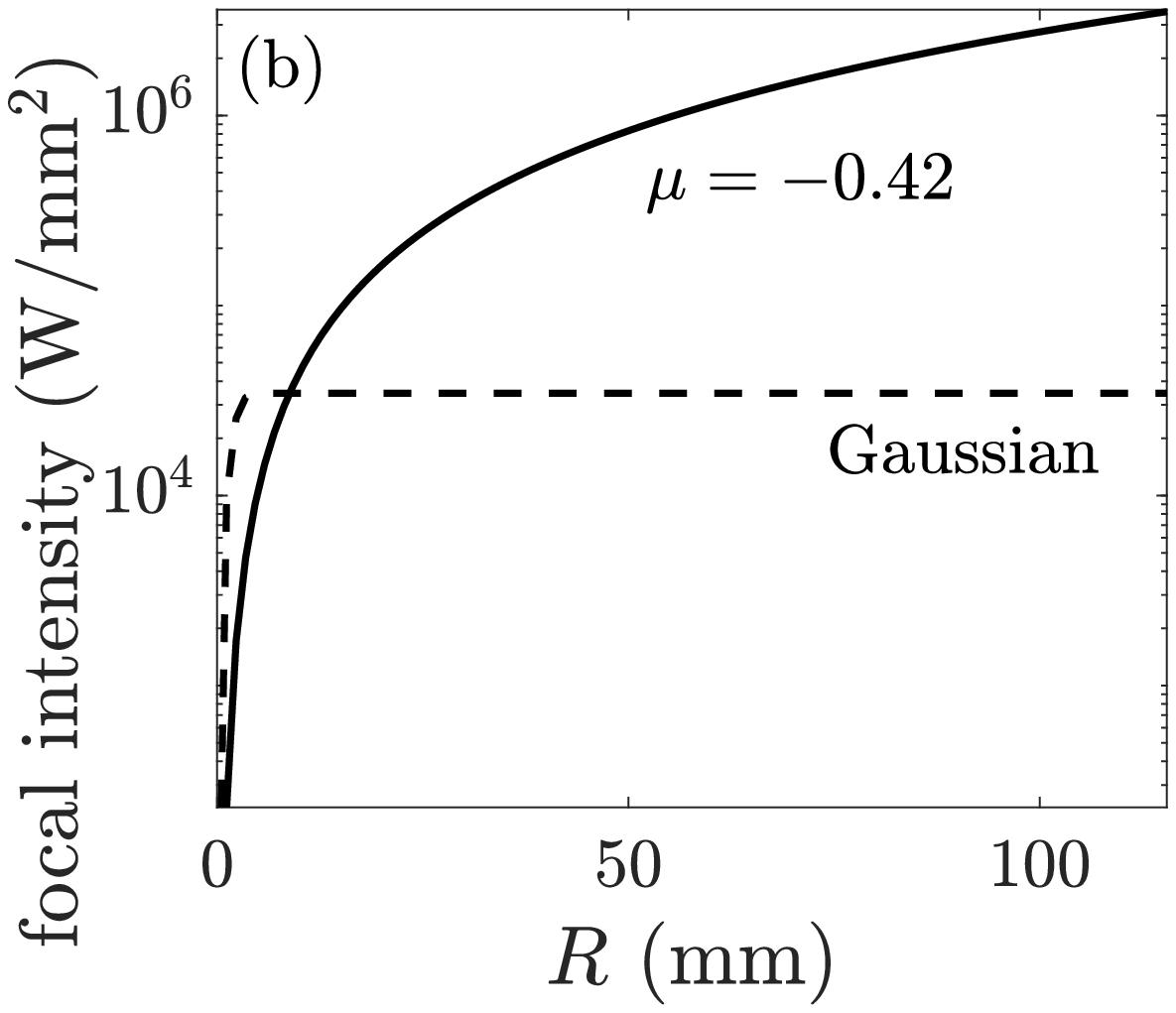}
\includegraphics*[height=3.75cm]{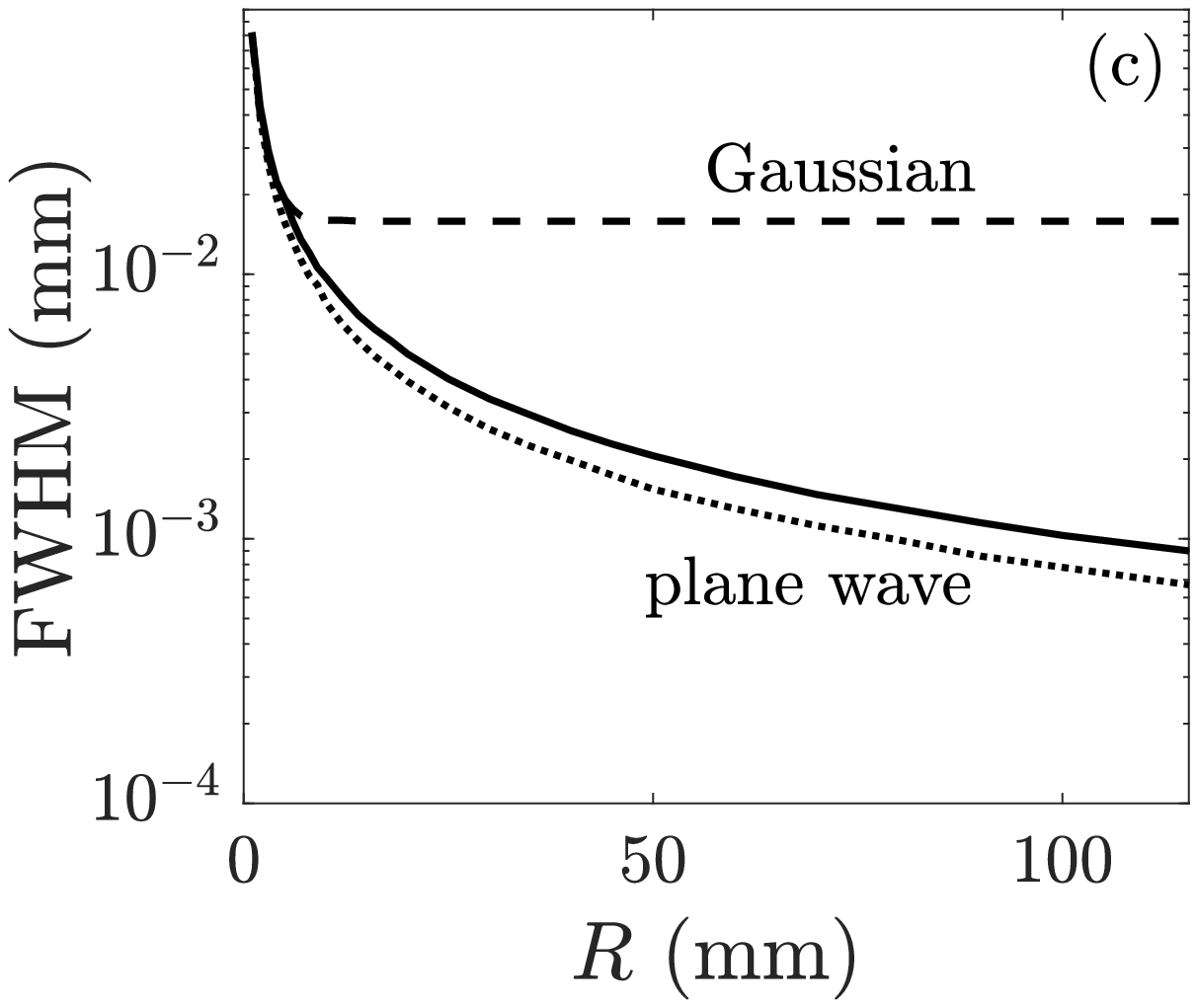}\includegraphics*[height=3.75cm]{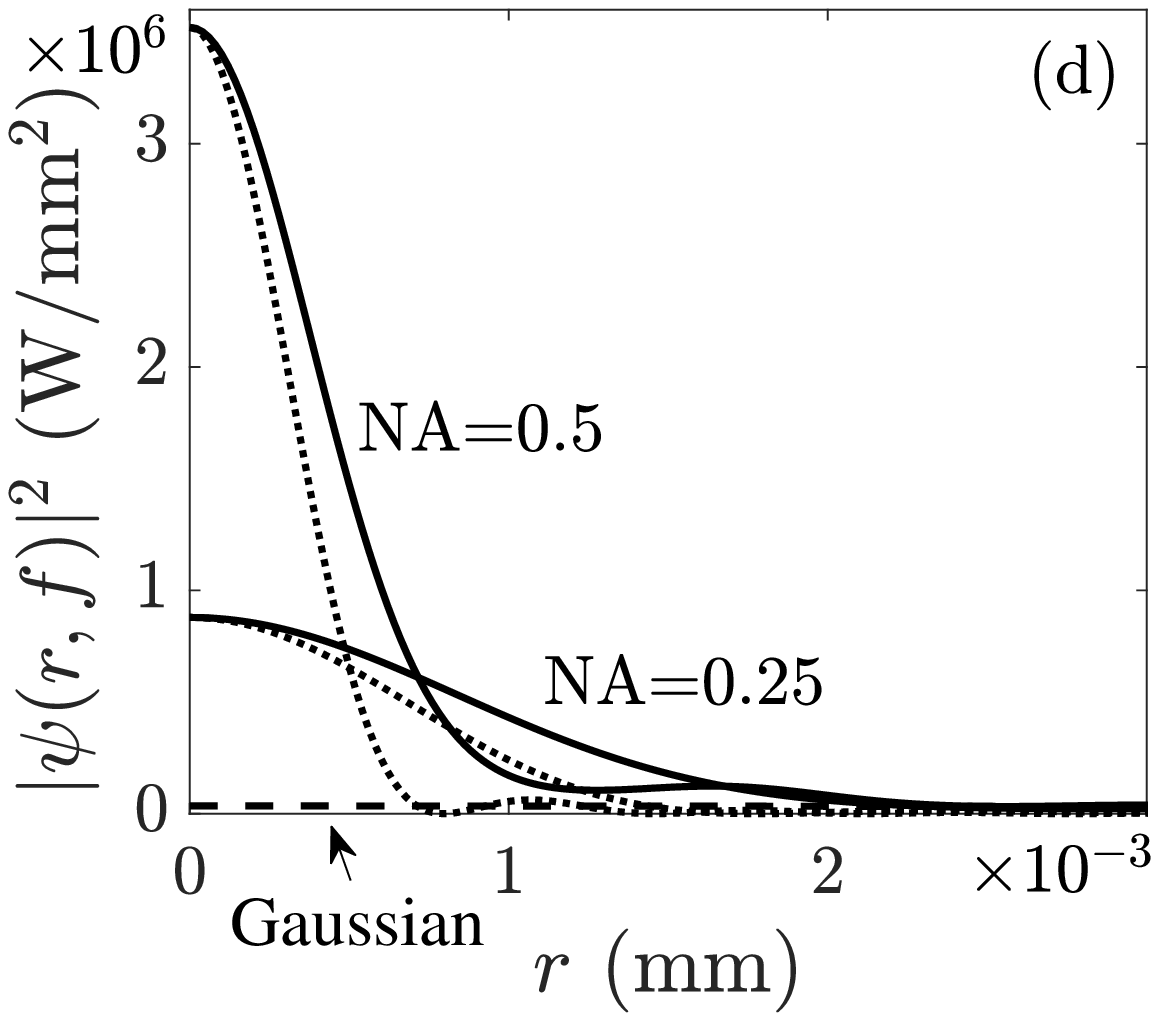}
\end{center}
\caption{\label{Fig3} {\it Physical properties of finite-aperture exploding beams.} (a) Focal intensity when exploding profiles (\ref{ILU}) with $s=0$, $\sigma=1$ mm, power $P=10$ W, frequency $\omega=2.5$ fs$^{-1}$ and the indicated values of $\mu$ illuminate a lens of focal length $f=200$ mm and increasing radius $R$. (b) The same in vertical logarithmic scale to visualize the much smaller focal intensity produced by Gaussian illumination of the same power and peak intensity on the lens. (c) FWHM diameter of the focal spot produced by the exploding profile (\ref{ILU}), by the Gaussian profile, and by a plane wave, as functions of the radius of the lens. (d) Radial profiles of intensity at the focal plane due to exploding (solid), Gaussian of the same peak intensity and power (dashed), and plane wave (dotted) illuminations for lens radii $R=51.65$ mm (0.25 NA) and $R=115.47$ mm (0.5 NA). For the Gaussian illumination there is no change with $R$.}
\end{figure*}

The above description is an ideal behavior that originates from ignoring the finite aperture of focusing systems, and that disappears when finite transversal extents are considered. Still, the existence of these singular beams in a linear medium has observable effects in experiments that are absent with standard illuminating beams. If a Gaussian or any other standard light beam carrying finite power is focused, the width of the focal spot and its intensity do not change once the lens aperture is appreciably larger than the beam spot size; by contrast, focusing the exploding  field in (\ref{ILU}) produces, in principle, a brighter and narrower focal spot as the aperture radius $R$ of the focusing system is opened. This behavior also mimics what happens when a plane wave illuminates the lens of increasing radius $R$, with the important difference that the focused power does not change significantly when opening the aperture (once $R\gg \sigma$), since the power in the beam tails is increasingly negligible. In addition, focusing the exploding vortex fields in (\ref{ILU}) produces increasingly brighter and narrower ring surrounding the vortex at the focal plane when opening the aperture without significantly changing the beam power. These effects are described below under conditions of paraxial focusing, conditions under which they are supposed to be physically valid.

%With tighter, nonparaxial focusing, these peculiar effects cease substantially, as described in the Appendix.

For beams without orbital angular momentum ($s=0$, hence $-1/2 < \mu <0$), the intensity at the focal point $(r,z)=(0,f)$ with a lens of aperture radius $R$ can easily be evaluated from (\ref{FRESNELR}) with (\ref{ILU}) to be
\begin{equation}\label{FOCALR}
|\psi(0,f)|^2=\frac{P}{A}\left(\frac{k\sigma^2}{2 f|\mu|}\right)^2\left[1-\left(1+\rho^2\right)^{|\mu|}\right]^2\, ,
\end{equation}
where $\rho=R/\sigma$,
which is seen in Fig. \ref{Fig3}(a) to grow without bound with $R$ for any allowed value of $\mu$ even when the aperture radius $R$ is much larger than the beam spot size of (\ref{ILU}) on the lens, measured by $\sigma$. The curves end when the angle of a marginal ray as seen from the focus is $30^\circ$, taken, according to Siegman \cite{SIEGMAN}, as a limit of validity of the paraxial approximation, i. e., for a numerical aperture of NA $=0.5$. Evaluation of the derivative of (\ref{FOCALR}) with respect to $\mu$ reveals more pronounced intensity enhancements (at large values $R/\sigma$) when using values of $\mu$ about $-0.42$, as can also be appreciated in Fig. \ref{Fig3}(a). By contrast, the intensity at the focal point of Gaussian illumination approaches a constant value as soon as the aperture radius $R$ is slightly larger than its spot size, as seen in Fig. \ref{Fig3}(b). Note the logarithmic vertical scale to visualize the tiny focal intensity of the focused Gaussian illumination of the same power and peak intensity. Intensities two orders of magnitude higher than those attainable with the Gaussian illumination are obtained while keeping paraxial focusing conditions. Also, the diameter of the focal spot decreases as the aperture radius is increased up to the limit imposed by the paraxial approximation, in contrast with the constant and much larger focal diameter with the Gaussian beam of the same power and peak intensity, as seen in Fig. \ref{Fig3}(c). Bright spots of area two orders of magnitude smaller than with the Gaussian illumination are obtained.  Two examples of focused transversal profiles with numerical apertures $0.5$ and $0.25$ are compared to the low-intensity and much broader focused Gaussian beam in Fig. \ref{Fig3}(d).

Figures \ref{Fig3}(c) and (d) include also diffraction-limited focal diameters and transversal profiles (dotted curves) corresponding to plane wave illumination on the lens of radius $R$. It is clear that focusing the exploding profile with finite power imitates focusing of a plane wave, in the sense that increasing the numerical aperture results in tighter focal spots. The peak intensities of the focused transversal profiles with illumination (\ref{ILU}) and a plane wave are equated in Fig. \ref{Fig3}(d) to visualize that the width of the focal spot is, for any aperture radius, only slightly above the diffraction limit, in contrast to what happens with the Gaussian illumination.

\begin{figure*}[!]
\begin{center}
\includegraphics*[height=3.75cm]{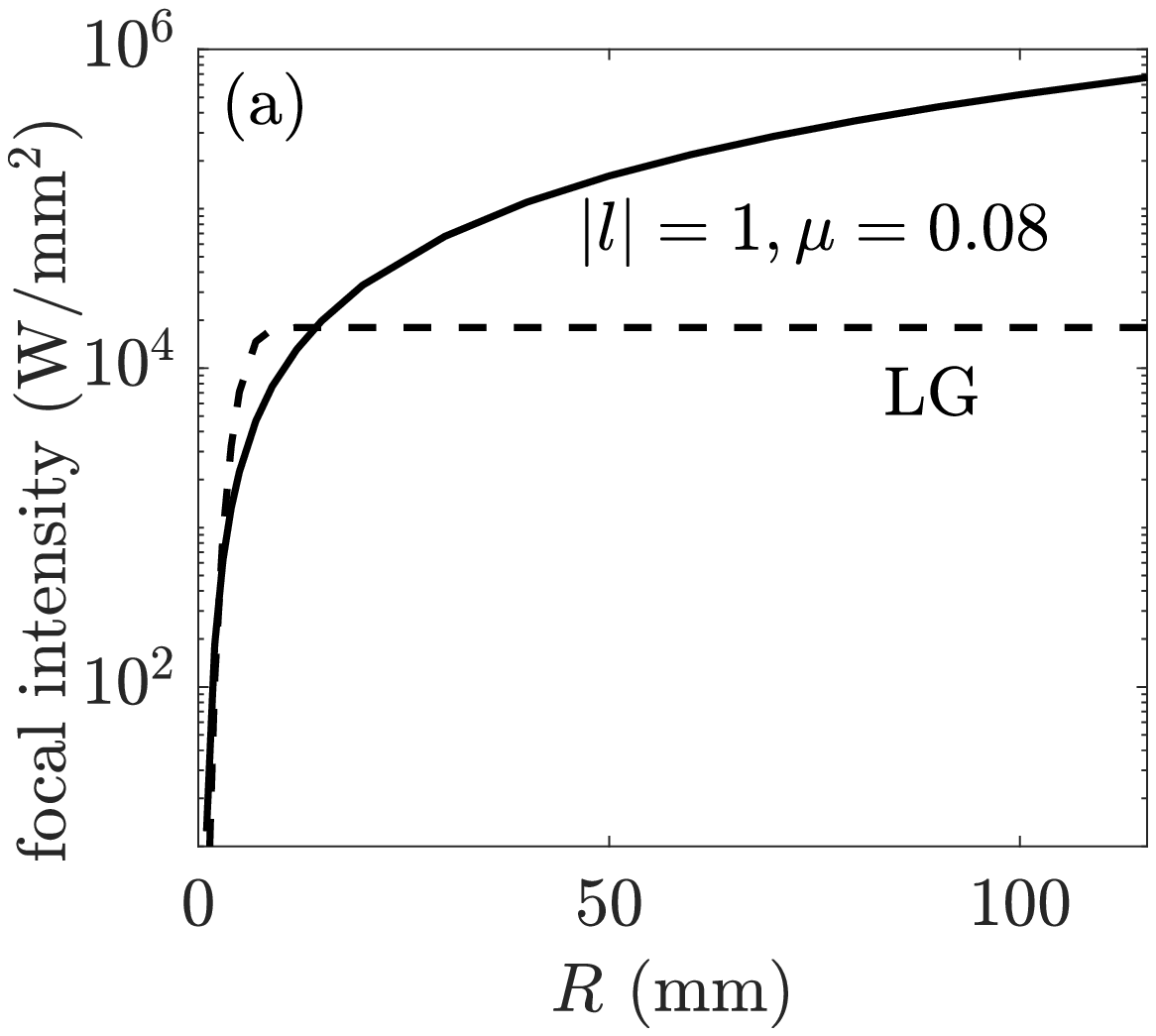}\includegraphics*[height=3.65cm]{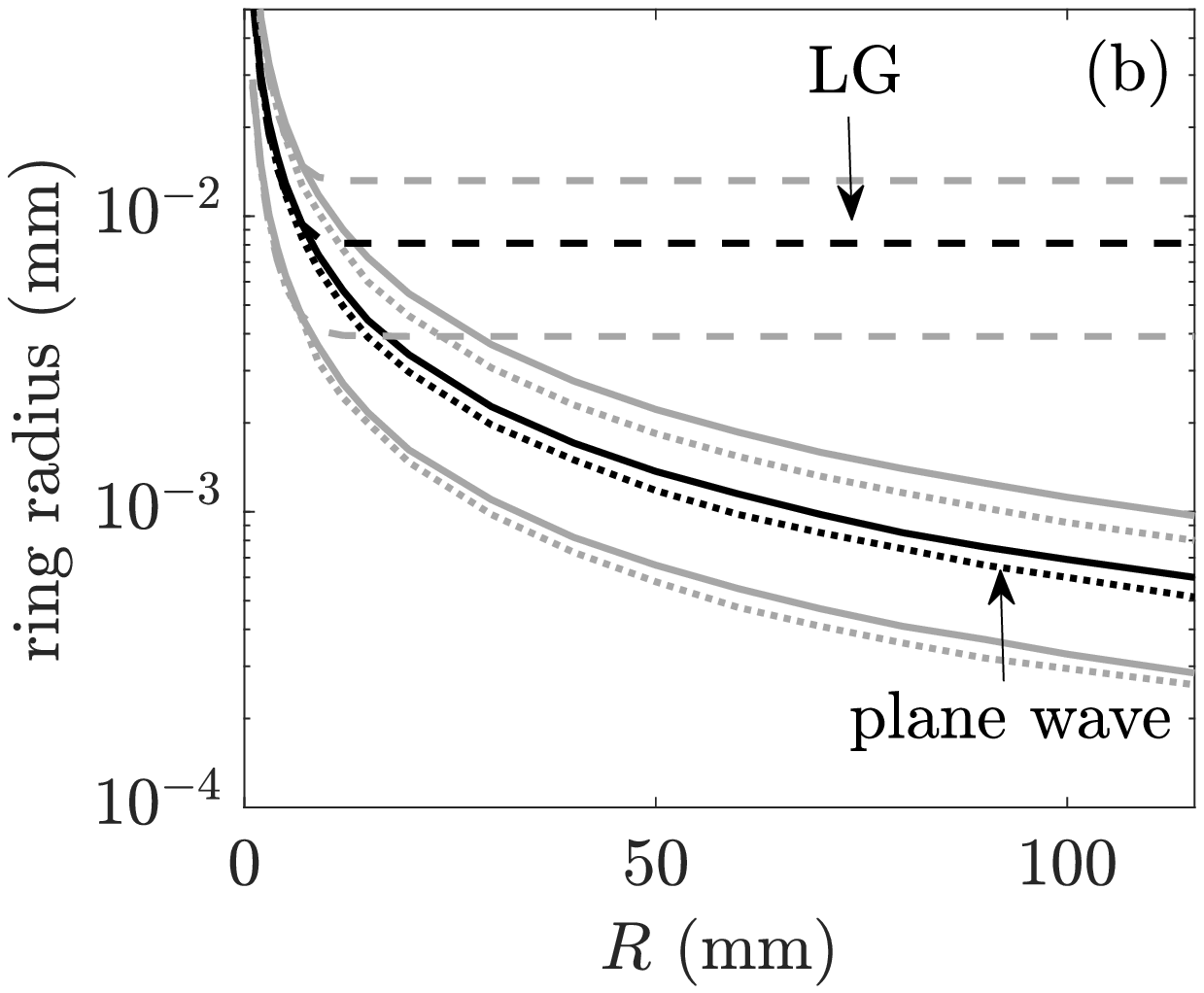}
\includegraphics*[height=3.65cm]{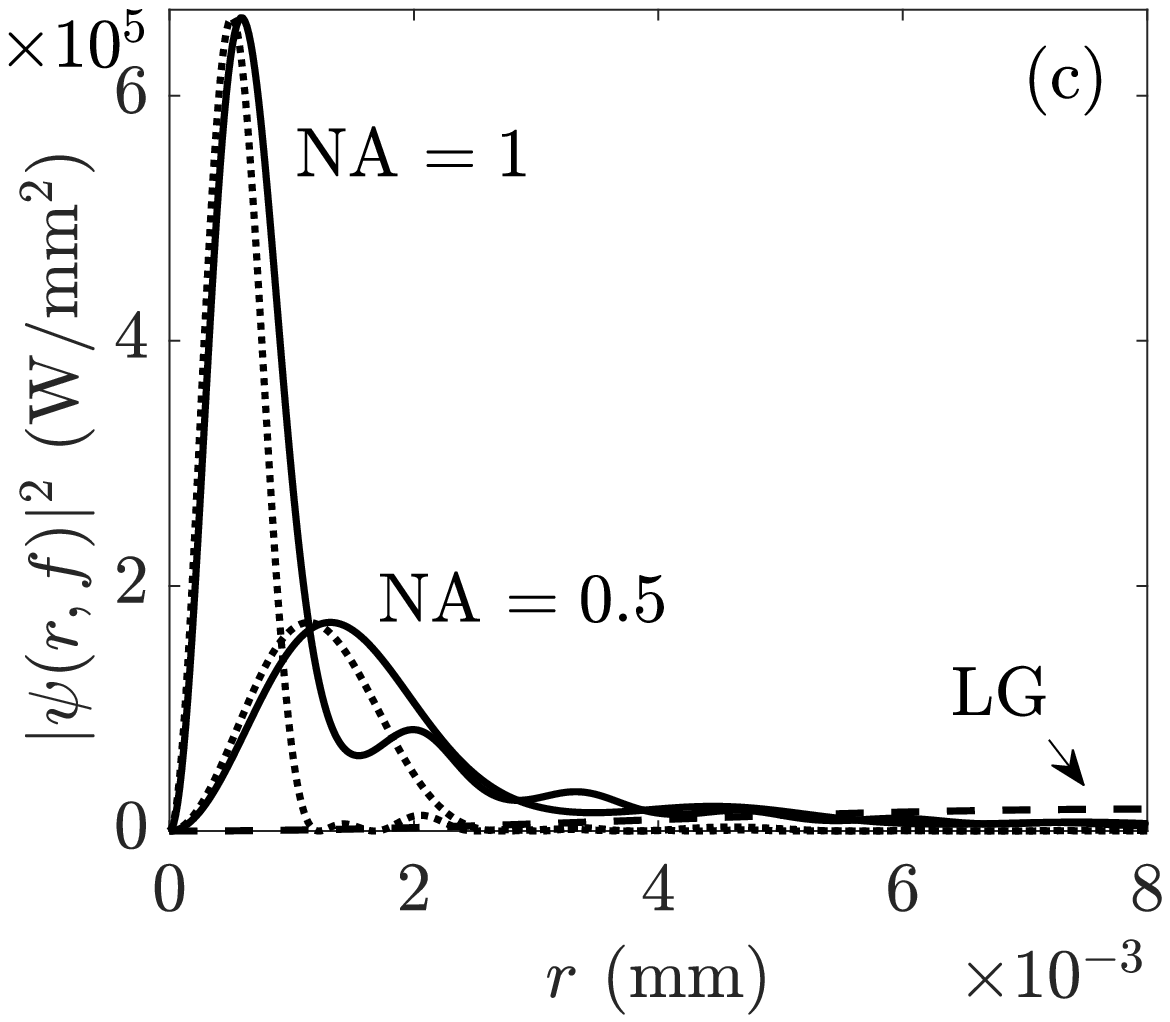}\includegraphics*[height=3.75cm]{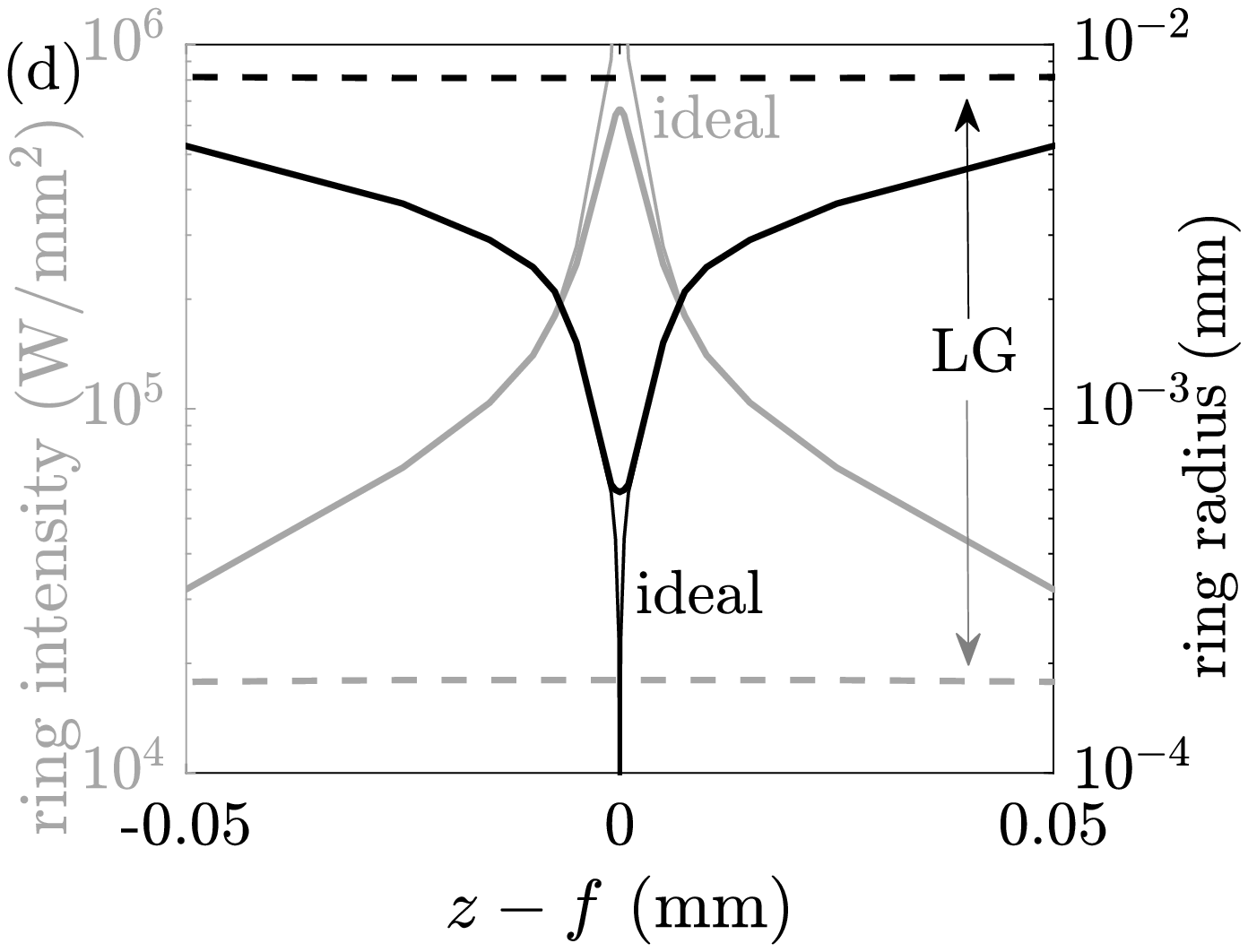}
\end{center}
\caption{\label{Fig4} {\it Physical properties of finite-aperture exploding vortex beams.} (a) Focal intensity when exploding profiles (\ref{ILU}) with $|s|=1$, $\sigma=1$ mm, power $P=10$ W, frequency $\omega=2.5$ fs$^{-1}$ and $\mu=0.08$ illuminate a lens of focal length $f=200$ mm and increasing radius $R$ (solid curve) compared to the same quantity for Laguerre-Gauss (LG) illumination of the same power and peak intensity on the lens. (b) Radius of the maximum of the bright ring at the focal plane  produced by the exploding profile (\ref{ILU}) (solid), by the Laguerre-Gauss profile (dashed), and by an uniform plane wave with a punctual vortex (dotted), as functions of the radius of the lens. The gray curves of the same type represent the inner radius and outer radius (half width at half maximum of the maximum intensity) for the corresponding illuminations. (c) Radial profiles of intensity at the focal plane due to exploding (solid), Laguerre-Gauss of the same peak intensity and power (dashed), and vortex plane wave (dotted) illuminations for lens radii $R=51.65$ mm (0.25 NA) and $R=115.47$ mm (0.5 NA). For the Gaussian illumination there is no change with $R$. (d) Intensity of the bright ring and its radius as a function of propagation distance about the focus for the exploding beam with radius $R=115.47$ mm (solid curves) compared to the ideal exploding beam. The axial region is so short that the same quantities for Laguerre-Gauss beams do not appreciably change (dashed lines).}
\end{figure*}

For exploding beams with orbital angular momentum ($|s|\neq 0$, hence $(|s|-1)/2< \mu <|s|/2$) the intensity of the bright ring at the focal plane grows similarly when opening the aperture, in contrast to what happens to Laguerre-Gauss beams, as seen in Fig. \ref{Fig4}(a). As with $s=0$, this enhancement is further magnified by using the value $\mu \simeq (|s|-1)/2+0.08$, e. g., the value $\mu = 0.08$ for $|s|=1$ chosen in Fig. \ref{Fig4}. Again focal intensities about two orders of magnitude higher than with Laguerre-Gaussian illumination of the same power, peak intensity and vorticity on the lens are easily attained. Figure \ref{Fig4}(b) shows the shrinking radius of the bright ring at the focal plane when increasing the aperture radius $R$, in contrast to the constant radius with the Laguerre-Gauss illumination, and mimicking the shrinking diffraction-limited radius for plane wave illumination with a punctual vortex (solid, dashed and dotted curves, respectively). The gray curves of the same type locate the inner and outer radius of the corresponding bright rings, in order to visualize the constant thickness of the bright ring at the focal plane for the Laguerre-Gauss illumination, and the shrinking thicknesses for exploding and plane wave illuminations. Two examples of transversal profiles at the focal plane with numerical apertures $0.5$ and $0.25$ are seen in Fig. \ref{Fig4}(c), where they are compared to the much broader and less intense focused Laguerre-Gauss, and are also seen to be slightly broader than the diffraction-limited profiles of the same numerical apertures. Interestingly, the exploding beam focused by a lens of finite radius behaves very approximately as the ideal exploding beam from an infinite aperture out of the focal plane, as illustrated in Fig. \ref{Fig4}(d) plotting the intensity of the maximum of the bright ring and its radius about the focus. Only in a tiny fraction (about one wave length) of the focal region (of about $0.5$ mm) the bright ring intensity and its radius depart from the ideal behavior, reaching at the focus finite and non-zero values, respectively.

\section{Enhanced longitudinal component}

The ideally infinite gradients of these singular beams and vortex beams at the focal plane suggests the presence of a strong longitudinal or axial component of the electric field, even under paraxial conditions, since the axial component is directly related to the gradient of the transversal part. Enhancement of the axial component commonly relies on nonparaxial focusing and the use of specific fields such as radially polarized beams \cite{QUABIS,COLLIN}. While paraxial, standard, radially polarized fields present quite small longitudinal components, exploding radially polarized fields have longitudinal components comparable in amplitude to that of the transversal component.

For a paraxial beam of transversal components $\boldsymbol{\psi}_\perp =\psi_x \boldsymbol{u}_x + \psi_y\boldsymbol{u}_y$, the axial component can be evaluated from \cite{LAX}
\begin{equation}\label{LONG}
\psi_z = \frac{i}{k} \nabla_\perp \cdot \boldsymbol{\psi}_\perp = \frac{i}{k}\left(\frac{\partial \psi_x}{\partial x} + \frac{\partial \psi_y}{\partial y}\right)\,.
\end{equation}
Strictly speaking, and according to Gauss law for the electric field, the term $(i/k)\partial\psi_z/\partial z$ should be included in the right hand side of (\ref{LONG}), but according to the perturbative method in \cite{LAX}, the $x$ and $y$ derivatives are leading-order terms determining the axial component under paraxial conditions, and the $z$ derivative is a small correction. In addition, the axial component is evaluated below only at the focal plane, where it takes maximum values and therefore $\partial\psi_z/\partial z=0$.

In order to maintain the cylindrical symmetry of the intensity of the vectorial field, we consider left or right handed circular polarizations, $\boldsymbol{\psi}_\perp = \psi(r,z)e^{is\varphi} \boldsymbol{u}_{l,r}$, where $\boldsymbol{u}_{l,r} = (\boldsymbol{u}_{x}\pm i\boldsymbol{u}_{y})/\sqrt{2}$, and the upper and lower signs stand for left and right handed polarizations, respectively, and where $\psi(r,z)$ is given by Fresnel integral in (\ref{FRESNELR}) with (\ref{ILU}). Some calculation of derivatives leads to
\begin{equation}
\psi_z(r,z)e^{i(s\pm 1)\varphi} = \frac{1}{\sqrt{2}}\frac{i}{k}\left[\frac{\partial\psi(r,z)}{\partial r} \mp \frac{s}{r}\psi(r,z)\right]e^{i(s\pm 1)\varphi}\,.
\end{equation}
With $s=-1$ (hence $0<\mu <1/2$) and left handed circular polarization, and with $s=+1$ (also $0<\mu <1/2$) and right handed polarization, the axial components do not carry any vorticity, can then have maxima at the beam center (where the transversal components vanish), which are equal in both cases and given by $\psi_z(r,z)=(i/\sqrt{2}k)\left[\partial\psi(r,z)/\partial r + \psi(r,z)/r\right]$.

\begin{figure}[!]
\begin{center}
\includegraphics*[height=3.6cm]{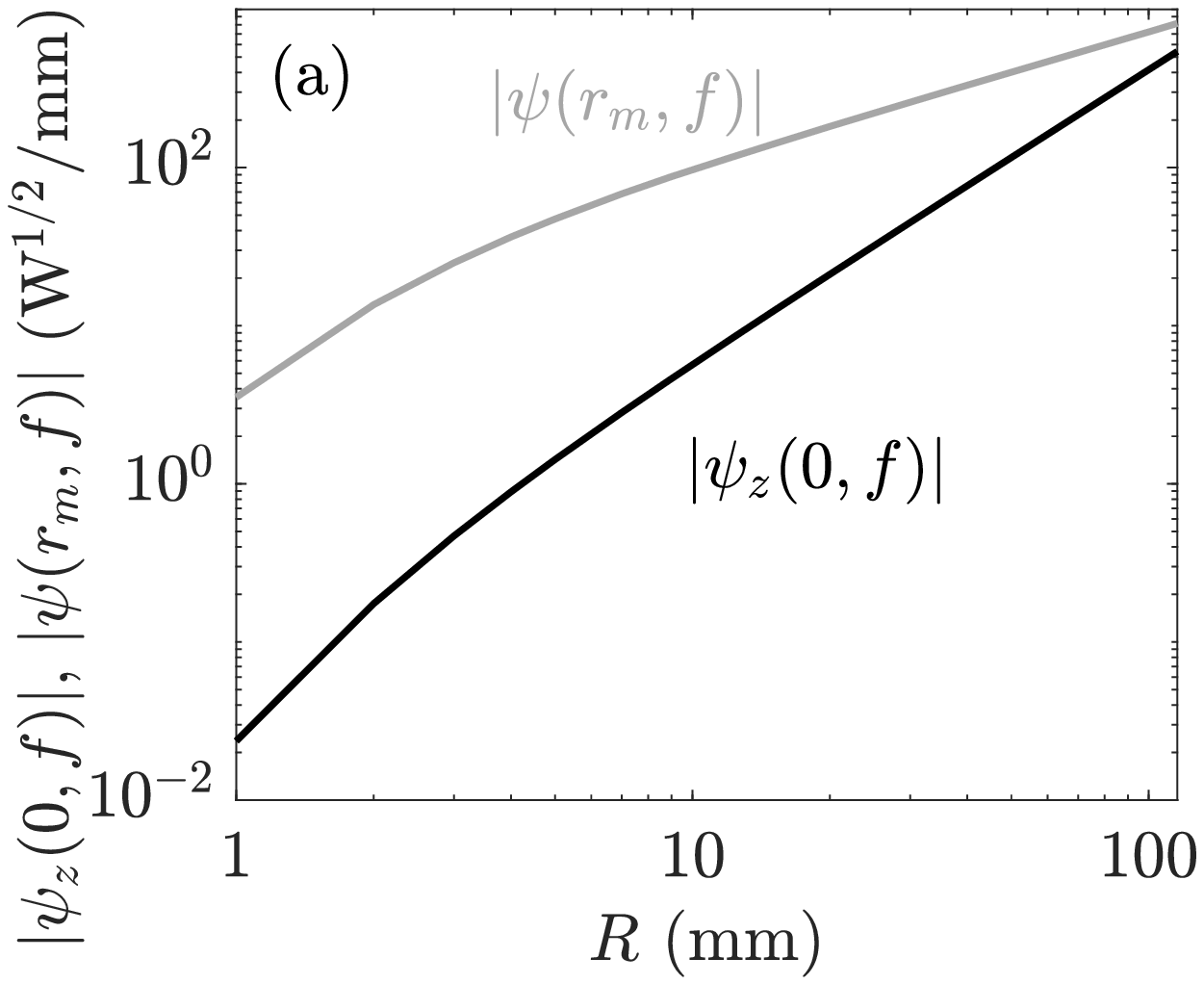}\includegraphics*[height=3.6cm]{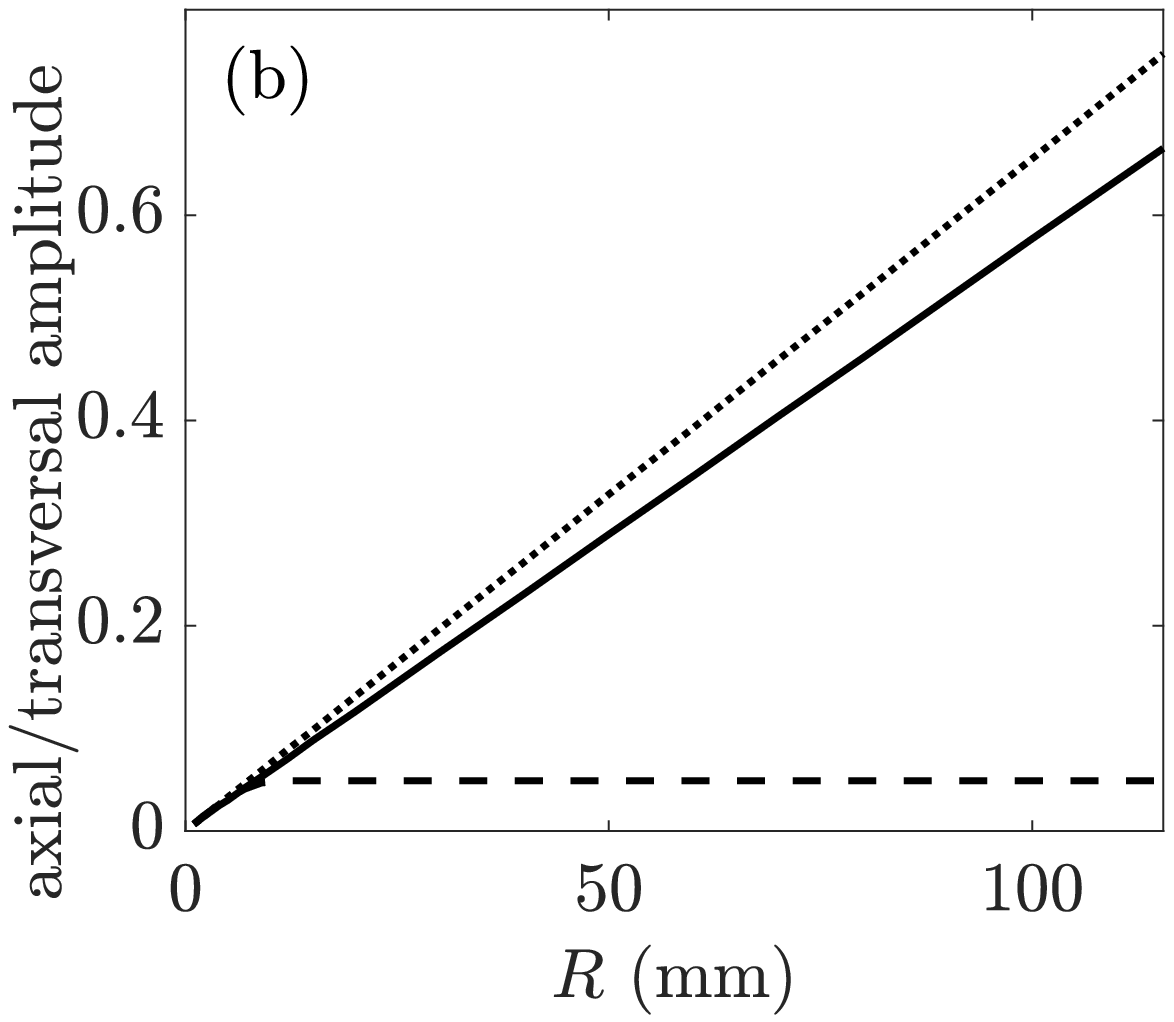}
\includegraphics*[height=3.6cm]{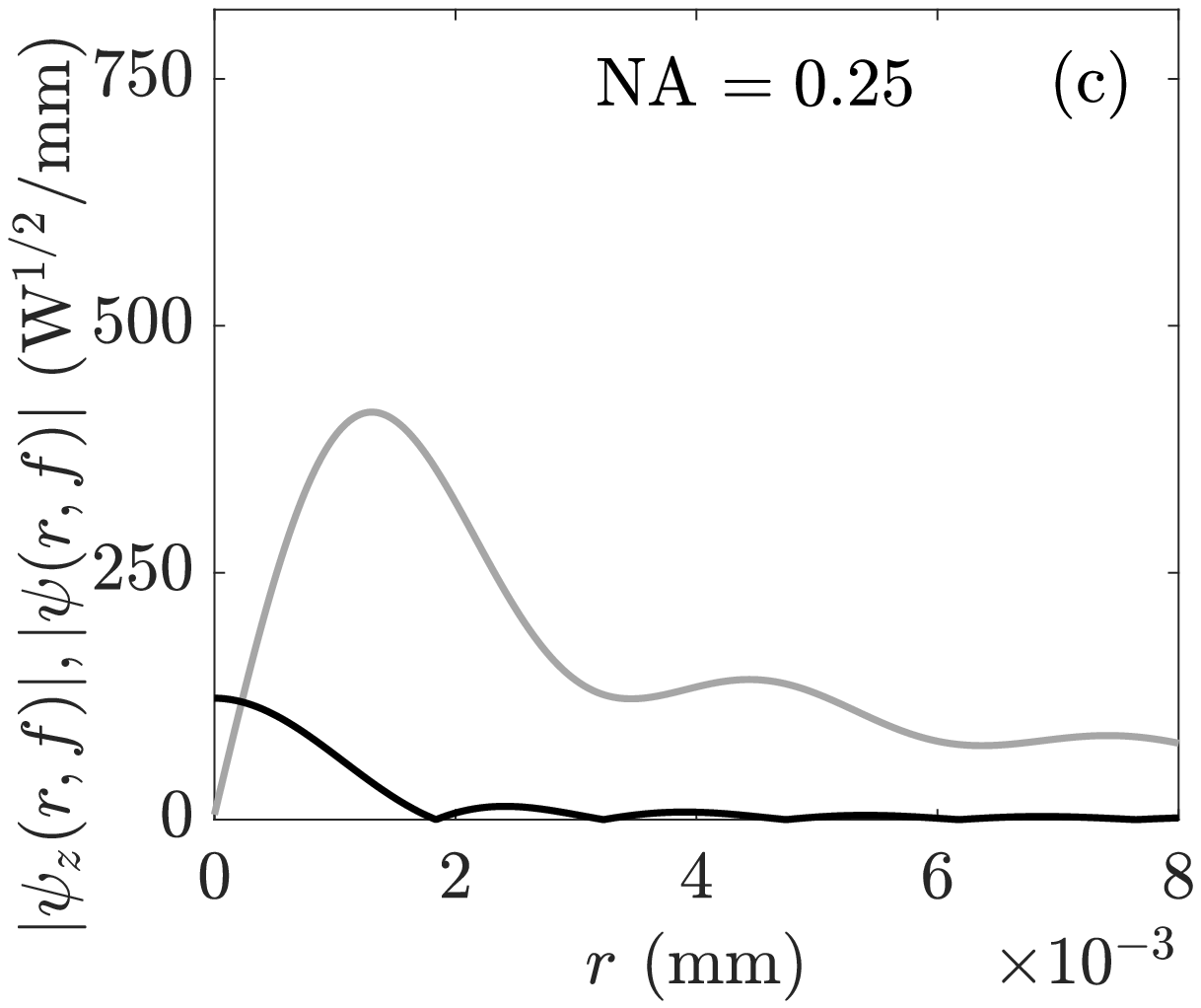}\includegraphics*[height=3.6cm]{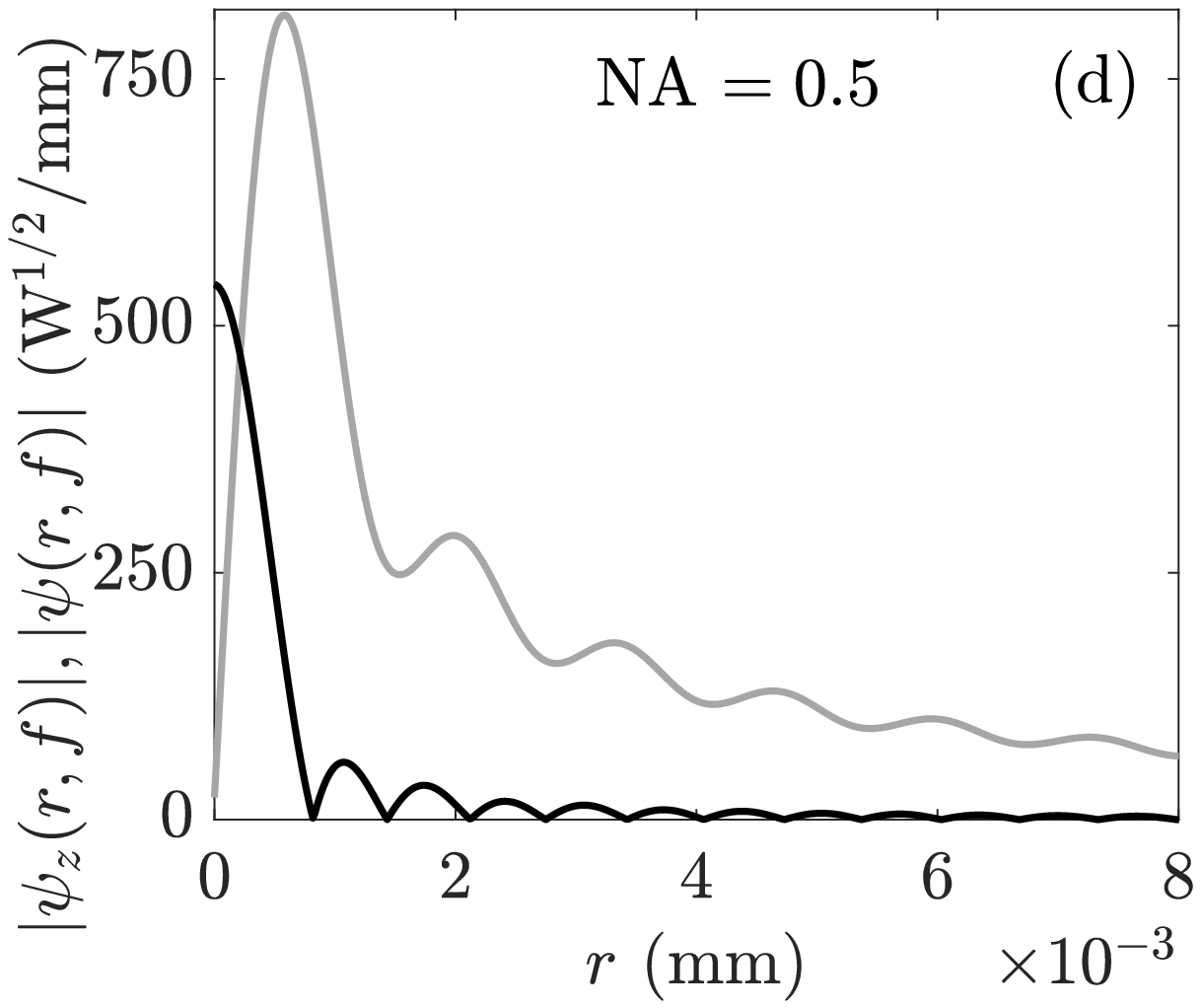}
\end{center}
\caption{\label{Fig5} {\it Exploding radially polarized beams.} (a) Amplitude at the focal point of the axial component $|\psi_z(0,f)|$ (black curve), as given by (\ref{AXIALF}), and of the transversal component $|\boldsymbol{\psi}_\perp(r_m,f)|=|\psi(r_m,f)|$ at its radial maximum $r_m$ (gray curve), when the radially polarized beam in (\ref{RAD}), with the exploding profile $\psi(r)$ in (\ref{ILU}) with $\sigma=1$ mm, $P=10$ W, frequency $\omega=2.5$ fs$^{-1}$, $\mu=0.08$ and $|s|=1$, is focused in vacuum with a lens of focal length $f=200$ mm, both as functions of the radius $R$ of the lens aperture. (b) Quotient of the axial and transversal amplitudes as a function of $R$ (solid curve) for the above conditions, for radially polarized Laguerre-Gauss illumination of the same power and peak intensity on the lens (dashed curve), and for radially polarized plane wave illumination with a punctual polarization singularity in its center (dotted curve). (c) and (d) Radial profiles of the amplitude of the axial (black) and transversal (gray) components for $R=51.65$ mm (NA= $0.25$) and $R=115.47$ mm (NA $=0.5$).}
\end{figure}

We can now easily consider the radially or azimuthally polarized beams
\begin{equation}\label{RAD}
\boldsymbol{\psi}_\perp = \frac{\psi(r,z)}{\sqrt{2}} \left[e^{-i\varphi}\boldsymbol{u}_l \pm e^{i\varphi}\boldsymbol{u}_r\right]\,,
\end{equation}
having the same amplitude associated with the transversal components, $|\boldsymbol{\psi}_\perp| = |\psi(r,z)|$, as that of the individual vortex beams with circular polarizations. The axial component vanishes for azimuthal polarization [for the minus sign in \ref{RAD}], and is given by
\begin{equation}\label{AXIAL}
\psi_z = \frac{i}{k}\left[\frac{\partial\psi(r,z)}{\partial r} + \frac{1}{r}\psi(r,z)\right] = \frac{i}{k}\frac{1}{r}\frac{\partial [r\psi(r,z)]}{\partial r}\,,
\end{equation}
for radial polarization [for the plus sign in \ref{RAD}]. Writing the ideal focal profile in (\ref{FOCAL}) for $|s|=1$ in the compact form $\psi (r,f)=C\alpha^\mu K_{1-\mu}(\alpha)$, where $\alpha=k\sigma r/f$ and $C$ is the constant in the first row in (\ref{FOCAL}), and where we have neglected the negligible curvature factor $e^{ikr^2/2f}$ in the Debye approximation (symmetric focused field about the focal plane), one can readily evaluate the axial component in (\ref{AXIAL}) as
\begin{equation}
\psi_z= \frac{i\sigma}{f} C\alpha^\mu \left[ K_{2-\mu}(\alpha) - 2 \frac{K_{1-\mu}(\alpha)}{\alpha}\right]\,,
\end{equation}
which is singular at $\alpha=0$, i. e., at $r=0$. Indeed the ratio of amplitudes of the axial and transversal components close to the origin behaves as $|\psi_z|/|\psi|= 2\sigma\mu/f \alpha = 2\mu/kr$, implying that the singularity in the axial field is stronger than that of the transversal field. Of course, these are mathematical equations for ideal exploding radially polarized beams, but they have physical manifestations in their apertured versions. From (\ref{AXIAL}), (\ref{FRESNELR}) with finite radius $R$ and illumination (\ref{ILU}) with $|s|=1$, the axial component of the radially polarized beam at the focal plane can be evaluated to be
\begin{equation}\label{AXIALF}
\psi_z(0,f) = \frac{ik}{f^2}\sqrt{\frac{P}{A}}\frac{\sigma^3}{2\mu}\left[\frac{\rho^2}{(1+\rho^2)^\mu} +
\frac{1-(1+\rho^2)^{1-\mu}}{1-\mu}\right] \,,
\end{equation}
where again $\rho=R/\sigma$. The amplitude of the axial component grows as $(R/\sigma)^{2-2\mu}$ at large $R/\sigma$, while the amplitude of the transversal component at the bright ring surrounding the polarization singularity grows with the potential law $(R/\sigma)^{1-2\mu}$ (also at large $R/\sigma$) of lower power, as can be seen in Fig. \ref{Fig5}(a). This difference implies a linear increase with $R/\sigma$ of the ratio of the axial and transversal amplitudes, as can be appreciated in Fig. \ref{Fig5}(b), that reaches a maximum value of $0.665$ in the limit of validity of the paraxial approximation (solid curve). As expected, for Laguerre-Gauss radially polarized beams, this ratio reaches a constant maximum value that does not exceed $0.049$ (dashed curve), and, again, the behavior of the axial component of the exploding radially polarized beam mimics the behavior of radially polarized uniform plane wave with a punctual polarization singularity in its center, whose maximum ratio is about $0.757$ (dotted curve). For illustration Figs. \ref{Fig5}(c) and (d) depict transversal profiles of amplitude of the axial and transversal components at the focal plane for NA $=0.25$ and $0.5$. The above analysis indicates that the individual, circularly polarized exploding vortex beams have also enhanced axial components, but we have ignored them because they simply are $\sqrt{2}$ smaller than that of the radially polarized beam.

\section{Exploding versus uniform illumination}

\begin{figure}[t]
\begin{center}
\includegraphics*[height=3.6cm]{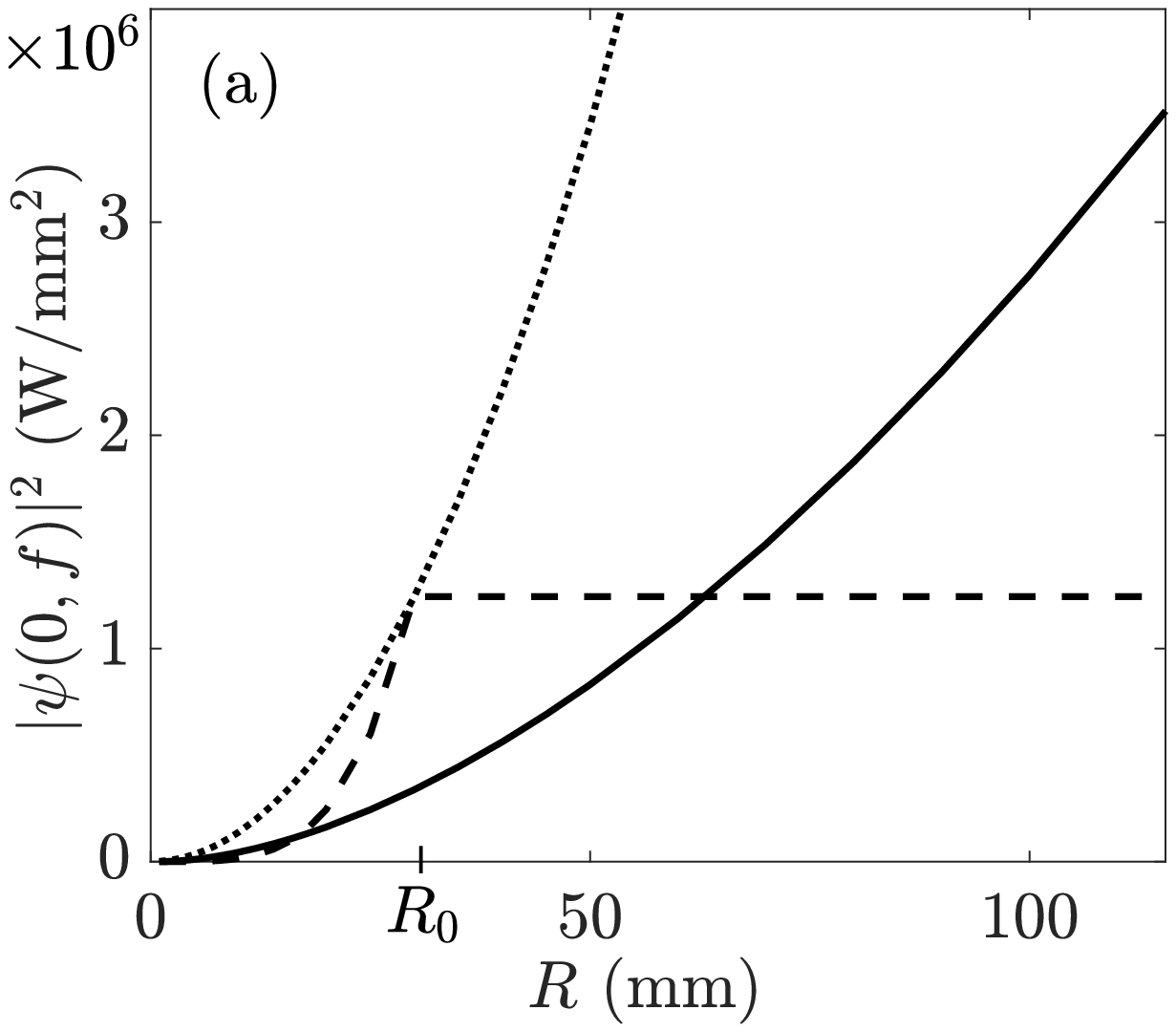}\includegraphics*[height=3.6cm]{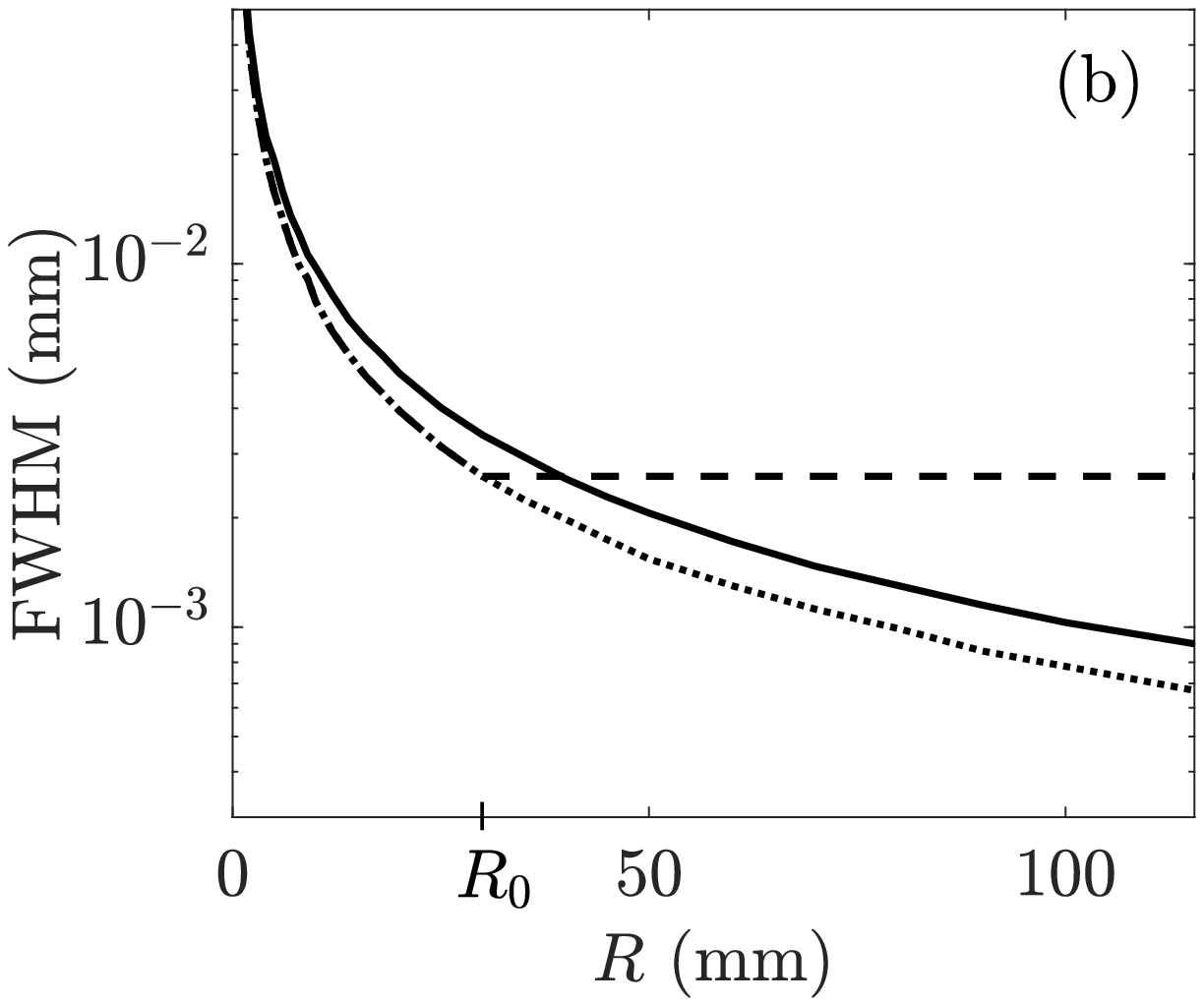}
\end{center}
\caption{\label{Fig6} Solid curves: For the focusing conditions and exploding illumination as in Fig. \ref{Fig3}, (a) peak intensity at focus and (b) FWHM of the focal spot as functions of the lens aperture $R$. Dotted curves: The same but for top-hat illumination of the same power of radius $R$, i. e., always filling the aperture. Dashed curves: The same for top-hat illumination of the same power and fixed radius $R_0= 30$ mm. }
\end{figure}

From what is exposed so far, it arises the practical question of the possible advantages of using exploding illumination over uniform illumination (of finite transversal extent). Of course the response depends on the particular application. Smooth profiles are more convenient in nonlinear optics applications to avoid instabilities and collapse. Reshaping Gaussian-like beam profiles into uniform flat-top profiles involves large power losses if based on truncation; otherwise it requires using sophisticated diffractive and refractive optics \cite{DICKEY}, or more recently developed techniques \cite{FRIES}, but uniform illumination provides better resolution \cite{MAZNEV} in applications such as microscopy.

Suppose here we dispose of the fixed amount of power $P$ supplied by a laser system which can be shaped as the exploding profile or as a flat-top profile, and wish to increase the intensity and/or diminish the size of the focal spot by opening the aperture, or simply to control these properties. As seen in Fig. \ref{Fig6} (a) and (b) the intensity is considerably higher and the spot size slightly smaller (the diffraction limit) with a flat-top profile that always fills the aperture (dotted curves) than with exploding illumination (solid curves), but this arrangement requires reshaping the flat-top profile each time the aperture radius is increased. The situation is different in a setting where reshaping of the illumination is difficult or not possible (dashed curves). The peak intensity of the flat-top illumination of the power $P$ and fixed radius $R_0$ is at first lower than with the exploding illumination because of truncation losses, then higher when more power is collected, and finally constant once $R_0$ is smaller than the aperture radius $R$, becoming eventually smaller than the focal intensity with the exploding illumination. Similarly, the focal spot is tighter but then constant and wider than with the exploding illumination of increasing aperture radius. Thus, the exploding profile offers the possibility of a smoother control of the intensity and width of the focused spot over wider ranges without the necessity of reshaping the input beam. Similar conclusion holds for the intensity and thickness of the bright ring of exploding vortex beams, and for the strength of the axial component of exploding radially polarized beams.

\section{Conclusions}

In conclusion, we have reported the existence of cylindrically symmetric, paraxial beams, vortex beams, and radially polarized beams of light that produce singular (infinite) intensities when they are ideally focused. They mimic the behavior of ideally focused plane waves but with a localized transversal profile that carries a finite amount of power.

With real focusing systems of finite lateral extent, this peculiar behavior manifests as focused beams whose intensity and spot size, vortex intensity and radius, and strength of the longitudinal component strongly change with the aperture size, even if the focused power is not significantly altered. With the same exploding illumination, focal intensities exceeding by two orders of magnitude, focal spots and vortex radii smaller by two orders of magnitude, and longitudinal component one order of magnitude higher than the same properties for standard illuminating beams of similar power and intensity can be achieved by increasing the aperture radius while keeping paraxial focusing.

These effects are absent with standard beams, and can find application in linear or nonlinear optics experiments where a precise control of the width and intensity of the focal spot, of the radius and intensity of the vortex bright ring, or the strength of the axial component of the electric field, are crucial, e. g., in second harmonic generation, also with radially polarized light \cite{YEW} and in surfaces \cite{BISS}, acceleration of electrons \cite{GUPTA,VARIN}, particle trapping \cite{ZHAN}, laser material processing \cite{DREVINSKAS}, etc. These exploding beams can also be used as alternative beams to standard Laguerre-Gauss beams for quantum entanglement of states of orbital angular momentum \cite{FICKLER}. Outside the field of optics, given the generality of the Schr\"odinger equation, exploding free electron wave packets \cite{BLIOKH} and vortex electron beams \cite{MCMORRAN} do exist and could find application in electron microscopy, as well as exploding acoustic vortices for Mie particle trapping \cite{MARZO}.

The author acknowledges support from Projects of the Spanish Ministerio de Econom\'{\i}a y Competitividad No. MTM2015-63914-P, and No. FIS2017-87360-P.

\end{document}